\documentclass[%
amsmath,
amssymb,
prl,
superscriptaddress,
twocolumn]{revtex4-2}

\usepackage[utf8]{inputenc}
\usepackage{CJKutf8}
\usepackage{amsfonts,accents,amsthm}
\usepackage{amsmath}
\usepackage{mathtools}
\usepackage[colorlinks=true,citecolor=blue,linkcolor=blue,urlcolor=blue]{hyperref}
\usepackage{xcolor}
\usepackage{bbold}
\usepackage{float}
\usepackage{braket}
\usepackage{orcidlink}

\usepackage[german,english]{babel}

\makeatletter
\def\bbl@set@language#1{%
	\edef\languagename{%
		\ifnum\escapechar=\expandafter`\string#1\@empty
		\else\string#1\@empty\fi}%
	\@ifundefined{babel@language@alias@\languagename}{}{%
		\edef\languagename{\@nameuse{babel@language@alias@\languagename}}%
	}%
	\select@language{\languagename}%
	\expandafter\ifx\csname date\languagename\endcsname\relax\else
	\if@filesw
	\protected@write\@auxout{}{\string\select@language{\languagename}}%
	\bbl@for\bbl@tempa\BabelContentsFiles{%
		\addtocontents{\bbl@tempa}{\xstring\select@language{\languagename}}}%
	\bbl@usehooks{write}{}%
	\fi
	\fi}
\newcommand{\DeclareLanguageAlias}[2]{%
	\global\@namedef{babel@language@alias@#1}{#2}%
}
\makeatother

\DeclareLanguageAlias{en}{english}
\DeclareLanguageAlias{English}{english}
\DeclareLanguageAlias{Englisch}{english}
\DeclareLanguageAlias{EN}{english}
\DeclareLanguageAlias{en-US}{english}
\DeclareLanguageAlias{de}{german}
\usepackage{graphicx}
\usepackage[export]{adjustbox}

\newcommand{\PRLSec}[1]{\textit{#1}.---}

\usepackage{cleveref}

\newcommand{\sbb}{\overline{SS}}
\newcommand{\ssb}{S\sbar}
\newcommand{\sbs}{\sbar{}S}
\newcommand{\sbar}{\overline{S}}

\newcommand{\ISR}[1]{\widetilde{\ham}_{#1}(\ev)}

\newcommand{\ham}{H}
\newcommand{\evecC}{\vecAlt{\Psi}}
\newcommand{\evecCComponent}[1]{\Psi_{#1}}

\newcommand{\ev}{E}

\newcommand{\hamEff}{\widetilde{\ham}}

\newcommand{\idMat}{I}

\newcommand{\jMat}{J}

\newcommand{\vecAlt}[1]{\mathbf{#1}}

\renewcommand{\d}{{\rm d}}

\newcommand{\SIFootnote}{See the Supplemental Material for more examples of latently symmetric setups, a proof that the Zak phase in latent SSH models is quantized, as well as details regarding the electric circuit part.}

\begin{document}
	\title{Latent Su–Schrieffer–Heeger models}

	\author{Malte Röntgen\,\orcidlink{0000-0001-7784-8104}}
     \altaffiliation[]{%
	These authors contributed equally to this work.
}%
 \email{Corresponding author, email: malte.rontgen@univ-lemans.fr}
	\affiliation{%
		Zentrum für optische Quantentechnologien, Universität Hamburg, Luruper Chaussee 149, 22761 Hamburg, Germany
	}%
\affiliation{%
	Laboratoire d’Acoustique de l’Université du Mans, Unite Mixte de Recherche 6613, Centre National de la Recherche Scientifique, Avenue O. Messiaen, F-72085 Le Mans Cedex 9, France
}%
     \author{Xuelong Chen (陈学龙)}
     \altaffiliation[]{%
	These authors contributed equally to this work.
}%
\affiliation{%
	EIT Institute for Advanced Study, Ningbo, China
}%
\begin{CJK}{UTF8}{gbsn}
	\author{Wenlong Gao (高文龙)\,\orcidlink{0000-0002-3446-7157}}
  \email{Corresponding author, email: wenlong.gao@eias.ac.cn}
\affiliation{%
EIT Institute for Advanced Study, Ningbo, China
}%

	\author{Maxim Pyzh\,\orcidlink{0000-0002-0079-1426}}%
\affiliation{%
	Zentrum für optische Quantentechnologien, Universität Hamburg, Luruper Chaussee 149, 22761 Hamburg, Germany
}%

\author{Peter Schmelcher\,\orcidlink{0000-0002-2637-0937}}
	\affiliation{%
		Zentrum für optische Quantentechnologien, Universität Hamburg, Luruper Chaussee 149, 22761 Hamburg, Germany
	}%
	\affiliation{%
		The Hamburg Centre for Ultrafast Imaging, Universität Hamburg, Luruper Chaussee 149, 22761 Hamburg, Germany
	}%

	\author{Vincent Pagneux\,\orcidlink{0000-0003-2019-823X}}%
\affiliation{%
	Laboratoire d’Acoustique de l’Université du Mans, Unite Mixte de Recherche 6613, Centre National de la Recherche Scientifique, Avenue O. Messiaen, F-72085 Le Mans Cedex 9, France
}%

	\author{Vassos Achilleos\,\orcidlink{0000-0001-7296-2100}}%
	\affiliation{%
		Laboratoire d’Acoustique de l’Université du Mans, Unite Mixte de Recherche 6613, Centre National de la Recherche Scientifique, Avenue O. Messiaen, F-72085 Le Mans Cedex 9, France
	}%

 	\author{Antonin Coutant\,\orcidlink{0000-0002-5526-9184}}%
   \email{Corresponding author, email: coutant@lma.cnrs-mrs.fr}
\affiliation{%
	Aix Marseille University, CNRS, Centrale Marseille, LMA UMR 7031, Marseille, France
}%

	\begin{abstract}
 The Su–Schrieffer–Heeger (SSH) chain is the reference model of a one-dimensional topological insulator.
 Its topological nature can be explained by the quantization of the Zak phase, due to reflection symmetry of the unit cell, or of the winding number, due to chiral symmetry.
 Here, we harness recent graph-theoretical results  to construct families of setups whose unit cell features neither of these symmetries, but instead a so-called latent or hidden reflection symmetry.
This causes the isospectral reduction---akin to an effective Hamiltonian---of the resulting lattice to have the form of an SSH model. As we show, these latent SSH models exhibit features such as multiple topological transitions and edge states, as well as a quantized Zak phase. Relying on a generally applicable discrete framework, we experimentally validate our findings using electric circuits.
	\end{abstract}
	
	\maketitle
 \end{CJK}
	
	\PRLSec{Introduction}
In the past years, topological insulators have become a main research focus of condensed matter and wave physics.
A prototypical model of a one-dimensional topological insulator is the Su–Schrieffer–Heeger (SSH) chain \cite{Su1979PRL421698SolitonsPolyacetylene}.
The SSH chain hosts topological edge states, which can be understood from the quantization of the Zak phase due to the unit cells mirror symmetry.
Alternatively, these topological states can be seen to be protected by chiral symmetry
and characterized by the winding number topological invariant \cite{Asboth2016ShortCourseTopologicalInsulators}.

The importance of the SSH chain as a simple and prototypical topological model is reflected in the large number of physical realizations, which include acoustic \cite{Chen2020PRA14024023ChiralSymmetryBreakingTightBinding,Coutant2021PRB103224309AcousticSuSchriefferHeegerLatticeDirect}, photonic \cite{Kremer2021OMEO111014TopologicalEffectsIntegratedPhotonic} or nanomechanical  \cite{Tian2019PRB100024310ObservationDynamicalPhaseTransitions} systems, as well as electric circuits \cite{Lee2018CP11TopolectricalCircuits}.
Furthermore, the original SSH model has been extended in different ways, for instance, by adding more sites to the unit cell \cite{Guo2015PRB91041402KaleidoscopeSymmetryprotectedTopologicalPhases,Anastasiadis2022PRB106085109BulkedgeCorrespondenceTrimerSuSchriefferHeeger,Xie2019nQI51TopologicalCharacterizationsExtendedSu}, by adding long-range couplings \cite{Perez-Gonzalez2019PRB99035146InterplayLongrangeHoppingDisorder}, or by converting the unit cell to a fractal \cite{Biswas2022FlatBandsEdgeStates}.

In this work, we propose a different class of extensions. Relying on recent advances in graph theory, we propose a class of models whose unit cells feature neither a chiral symmetry nor a reflection symmetry.
Instead, they are designed to posses a hidden (so-called latent \cite{Smith2019PA514855HiddenSymmetriesRealTheoretical}) symmetry.
In the past few years, latent symmetries have been used for a variety of purposes, including the generation of lattice systems with perfectly flat bands \cite{Morfonios2021PRB104035105FlatBandsLatentSymmetry, Leykam2018AP31473052ArtificialFlatBandSystems}, the explanation of accidental degeneracies in eigenvalue spectra \cite{Rontgen2021PRL126180601LatentSymmetryInducedDegeneracies}, or the design of asymmetric waveguide networks with broadband equireflectionality \cite{Rontgen2023PRL130077201HiddenSymmetriesAcousticWave,Sol2023CovertScatteringControlMetamaterials,Rontgen2023EquireflectionalityCustomizedUnbalancedCoherent}.
An introduction into latent symmetry is given in Ref. \cite{Rontgen2022LatentSymmetriesIntroduction}.
As we show, the effective Hamiltonian of models with latently symmetric unit cells are closely related to the classic SSH model.

In the following, we introduce the concept of such \emph{latent SSH} models, analyze their properties, and show how large families of them can be constructed.
Furthermore, we experimentally verify our predictions in electric circuit networks.

     \begin{figure}[htb] %[t]
	\centering
	\includegraphics[max width=\linewidth]{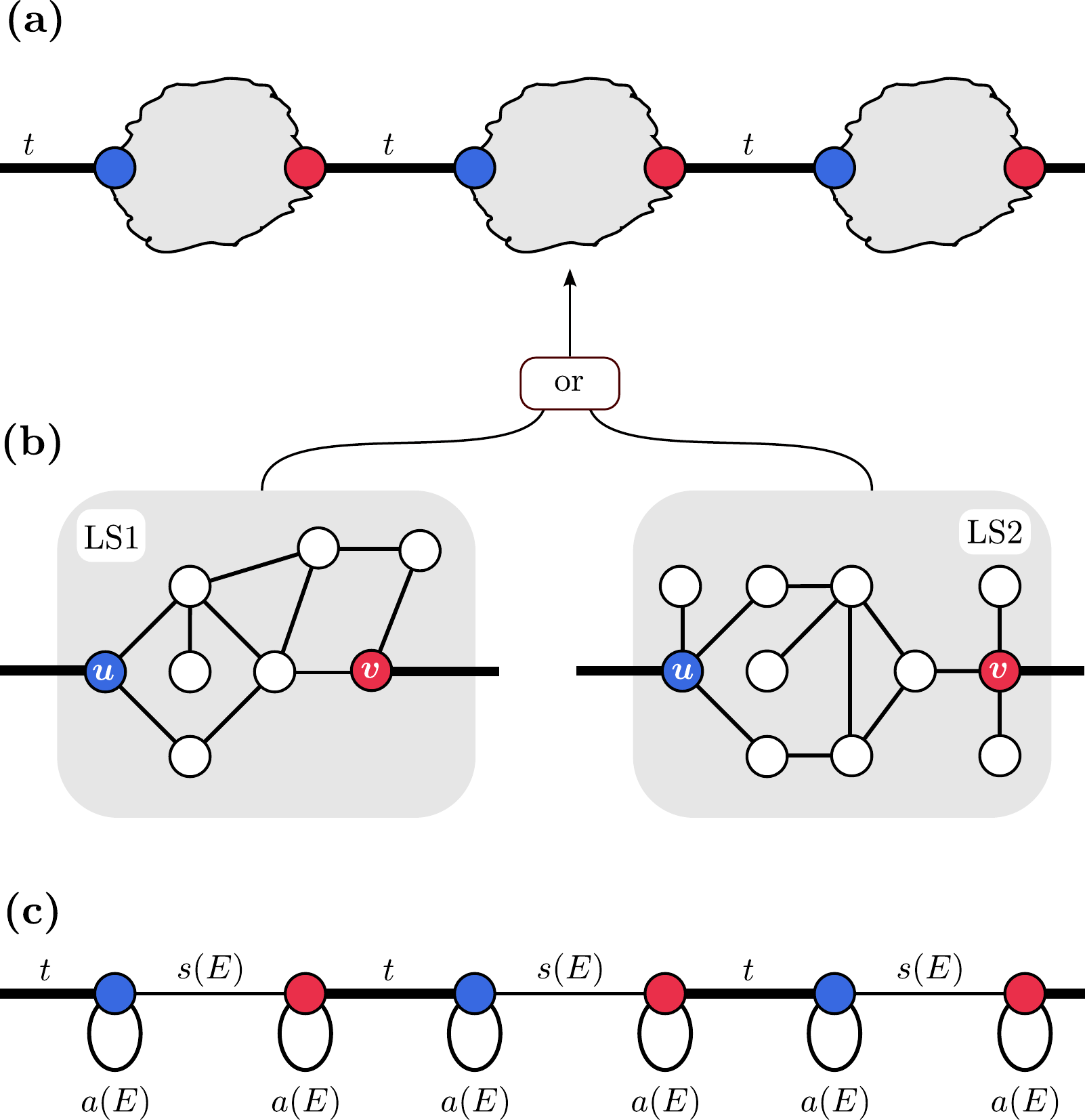}
	\caption{(a) The proposed one-dimensional lattice, with a cloud denoting a latently reflection symmetric subsystem. (b) Two unit cells LS1 and LS2. In both unit cells, the blue ($u$) and red ($v$) site are latently reflection symmetric. (c) Using a latently symmetric structure as a unit cell yields a latent SSH model. In all subfigures, each line corresponds to a coupling (with corresponding strength denoted, for instance, by $t$). Lines without annotations correspond to a coupling strength of unity. Loops denote on-site potentials.
	The sole purpose of (white, blue, red) colors is to highlight certain sites over others.
	}
	\label{fig:figure1}
\end{figure}
	
	\PRLSec{Topology through latent symmetry}
	Our main idea is to design topological lattices by relying on unit cells that feature neither a mirror symmetry nor a chiral symmetry, but a hidden mirror symmetry.
	Two such asymmetric, but hidden symmetric unit cells are depicted in \cref{fig:figure1} (b), while the overall structure of our proposed lattice is depicted in \cref{fig:figure1} (a).
	We emphasize that the unit cells shown in \cref{fig:figure1} (b) are only two out of many possible ones.
	We shall discuss the systematic generation of large families of such latently symmetric unit cells later on and show a couple of other examples in the Supplemental Material \setcounter{footnote}{16}\footnote{\SIFootnote}.
	
	Let us now build a lattice by choosing either LS1 or LS2 as a unit cell.
	We denote the Hamiltonian of our lattice by $H$; its eigenvalues and eigenstates can be obtained from solving the linear eigenvalue problem $\ham \evecC = E \, \evecC$.
	To uncover the hidden symmetry of the system, we will perform a so-called ``isospectral reduction'', which is akin to an effective Hamiltonian \cite{Grosso2013SolidStatePhysics}.
	To perform this reduction, one has to partition the system into two sets: $S$ and its complement $\sbar$. For our problem, we will choose $S$ to be all the blue and red sites in \cref{fig:figure1} (b), while $\sbar$ will be all the white sites.
	The isospectral reduction
	\begin{equation} \label{eq:ISRDef}
		\ISR{S} = \ham_{SS} + \ham_{\ssb} \left(\ev{}\, \idMat - \ham_{\sbb} \right)^{-1} \ham_{\sbs} \,,
	\end{equation}
	with  $\idMat$ the identity matrix, then converts our linear eigenvalue problem $\ham \evecC = E \, \evecC$ into the reduced, non-linear problem
	\begin{equation} \label{eq:reducedEVP}
		\ISR{S}\hspace{0.1em} \evecCComponent{S} = \ev{} \hspace{0.1em} \evecCComponent{S} \,.
	\end{equation}
	Here, $\evecCComponent{S}$ denotes the $S$ components of $\evecC$.
	We note that, in general, the original Hamiltonian and $\ISR{S}$ possess the same  eigenvalue spectrum
    \footnote{The eigenvalue spectrum of $\tilde{\ham}_S(E)$ is defined as the solutions to $Det(\tilde{\ham}_S(E) - E I )$.}, which motivates calling $\ISR{S}$ an isospectral reduction; see chapter 1 of Ref.  \cite{Bunimovich2014IsospectralTransformationsNewApproach}.
	The result of the above reduction on our lattice is shown in pictorial form in \cref{fig:figure1} (c).
	Taking the Fourier transform of the resulting system, \cref{eq:reducedEVP} is transformed to
	\begin{equation}
		\hamEff_{S}^{(B)}(E,k) \,  \evecCComponent{S}(k) = E \, \evecCComponent{S}(k)
	\end{equation}
	with the effective Bloch-Hamiltonian
		\begin{equation} \label{eq:latentBlochSSH}
				\hamEff_{S}^{(B)}(E,k) = \begin{pmatrix}
						a(\ev) & s(\ev) + t \, e^{i k} \\
						s(\ev) + t \, e^{-i k}& a(\ev)
					\end{pmatrix} \, .
			\end{equation}
			
	The crucial thing about $\hamEff_{S}^{(B)}$ is the equality of its two diagonal elements $a(E)$.	
	This equality is entirely due to a so-called latent reflection symmetry \cite{Smith2019PA514855HiddenSymmetriesRealTheoretical} of our unit cells LS1 and LS2.
	The name stems from the following observation: If one would take the isospectral reduction of the isolated unit cell (which is asymmetric) over $S=\{u,v\}$ (blue and red site), one would obtain a reflection-symmetric dimer.
	The isospectral reduction can thus be seen as a tool to uncover the hidden symmetry of the unit cell.

	Let us note that the equality of the diagonal elements in $\hamEff_{S}^{(B)}(E,k)$ would also occur if our unit cells would be reflection symmetric.
	For a generic unit cell which is neither reflection symmetric nor latently symmetric \footnote{Or when choosing $S$ such that it does not include sites which are related/mapped to each other by either a classical or a latent reflection symmetry.}, however, the two diagonal elements are generically unequal, though they may coincide for some specific values of $E$.

	Due to the equality of the diagonal elements, we can rewrite \cref{eq:latentBlochSSH} as
	\begin{equation} \label{eq:SSHAlgebra}
		\begin{pmatrix}
			0 & s(E) + t e^{i k} \\
			s(E) + t e^{-ik} & 0
		\end{pmatrix} \evecCComponent{S}(k) = \epsilon \, \evecCComponent{S}(k) \,,
	\end{equation}
	with $\epsilon = E - a(E)$.
	This takes formally the mathematical form of the classical SSH model, though with energy-dependent inter-cell coupling $s(E)$ and energy-dependent eigenvalue $\epsilon(E)$.
	We thus see that the isospectral reduction of our initial system mimics the SSH model; this justifies calling the initial system a latent SSH model.

    The algebraic mapping \cref{eq:SSHAlgebra} between the classic SSH model and our latent SSH model is the key to understand the properties of the latter.
    The original SSH model has a topological transition at eigenvalue zero when $s=t$.
    In \cref{eq:SSHAlgebra}, an eigenvalue $\epsilon = 0$ only occurs when $E = a(E)$.
    Consequently, each solution $E_i$ to $\epsilon(E) = 0$ corresponds to one topological transition, reached when the inter-cell coupling $|t| = |s(E_i)|$.
    Thus, the system is in the $i$-th topological phase when the inter-cell coupling $t$ fulfills $|t| > |s(E_i)|$.
    In a semi-infinite system, the $i$-th topological phase displays edge states whose amplitudes on the sites $u_n,v_n$ in the $n$-unit cell are $\evecCComponent{u_n} = \left( -\frac{s(E_i)}{t}\right)^n$, $\evecCComponent{v_n} = 0$ \footnote{This expression follows by analogy from the corresponding expression of the classic SSH model which can be found in chapter 1.5.6. of Ref. \cite{Asboth2016ShortCourseTopologicalInsulators}.}.
    We remark that the amplitudes on the remaining sites (the white ones, see \cref{fig:figure1} (b)) of each unit cell can be obtained through $\Psi_{\sbar} =  \left(\ham_{\sbb} - E \idMat \right)^{-1} \ham_{\sbs} \Psi_S$ and are, in particular, exponentially localized as well.

\begin{figure}[H] %[t]
	\centering
	\includegraphics[max width=\linewidth]{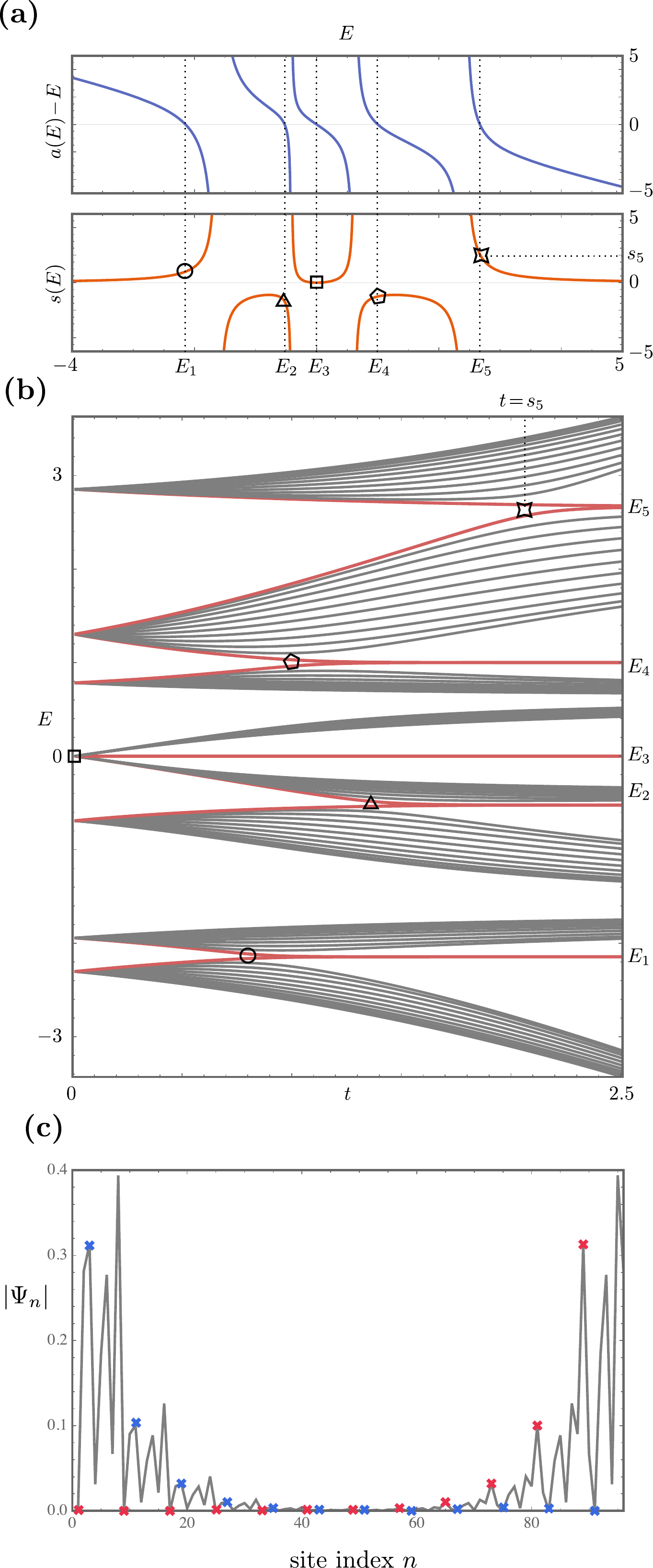}
	\caption{(a) The two curves $s(E)$ and $a(E) - E$ (see text for details), with the five topological transitions---occurring at energies $E_1$ to $E_5$---marked by different geometrical shapes. (b) The energy spectrum for a finite setup of $N=12$ unit cells in dependence of the inter-cell coupling $t$, with topological transitions marked by the same shapes as in (a). (c) The spatial profile of one of the two edge states at energy $E_1$ for $t=2.5$. Red/blue crosses denote the amplitude at sites $u,v$.
 For all cases, the lattice was built by using LS1 from \cref{fig:figure1} (b) as a unit cell.
	}
	\label{fig:figure2}
\end{figure}

    Let us now apply the above to a concrete setup.
    In \cref{fig:figure2}, we investigate the topological transitions for a setup with the $8$-site unit cell LS1 displayed in \cref{fig:figure1} (b).
    In particular, in \cref{fig:figure2} (a), we show the two curves $a(E) - E$ and $s(E)$.
	As can be seen, there are five topological transitions.
    In \cref{fig:figure2} (b), we show the energy eigenvalues for a finite setup of $N= 12$ unit cells (with $8$ sites per unit cell) for varying inter-cell coupling $t$.
    We clearly see that each transition is accompanied by the appearance of a pair of topological edge states that are 
    exponentially localized on the left or the right end of the chain, as is illustrated in \cref{fig:figure2} (c) for one of the states.

    Regarding the edge states of latent SSH models, two things are noteworthy.
    Firstly, by using the latent symmetry of sites $u,v$, the existence of these states can further be related to a quantized Zak phase of the unreduced system, as we show in the Supplemental Material \cite{Note17}.
    Secondly, it is well-known that the edge states of the conventional SSH lattice are robust with respect to disorder that respects chiral symmetry.
    Here, since the inter-cell coupling $t$ is independent of the energy, this robustness with respect to disorder in $t$ is---due to  \cref{eq:SSHAlgebra}---inherited by a latent SSH model as well.
	
	\PRLSec{Construction principles} A crucial ingredient to construct latent SSH models are latently symmetric unit cells, which can be easily designed using graph theoretical techniques \cite{Godsil1982AM25257ConstructingCospectralGraphs}.
    Alternatively, even when restricting oneself to systems with less than 12 sites and all unit couplings, one can easily generate several millions of latently symmetric setups through an exhaustive search.
    More specifically, using the nauty suite \cite{McKay2014JoSC6094PracticalGraphIsomorphismII}, one can efficiently construct all setups with a given number of sites, and then test each setup for a latent symmetry by scanning the first few matrix powers of each setup \cite{Rontgen2020PRA101042304DesigningPrettyGoodState}.
    We emphasize that latently symmetric systems are rather robust. That is, they can be altered in certain ways (changing certain couplings and/or on-site potentials) without breaking latent symmetry \cite{Rontgen2020PRA101042304DesigningPrettyGoodState,Morfonios2021LAaiA62453CospectralityPreservingGraphModifications}. For instance, as long as the on-site potentials of two latently symmetric sites remain identical, they can be set to any value. In the Supplemental Material, we show a couple of such changes of the unit cell LS2 while maintaining its latent symmetries \cite{Note17}.

    \PRLSec{Number of topological phases}
    An important property of a latent SSH model is that it can possess not only one but multiple topologically protected edge modes. The number of these modes is equivalent to the number of solutions to $a(E) = E$.
    Using the definition of the isospectral reduction, \cref{eq:ISRDef}, it can be shown that $a(E) = p(E)/q(E)$ is a rational function in $E$, with the degree $\mathrm{deg}(p) < \mathrm{deg}(q) \leq N-2$, where $N$ is the number of sites in the unit cell.
    The number of topological phases is then given by the degree of the numerator of $a(E) - E = \frac{p(E) - E q(E)}{q(E)}$.
    It follows that a latent SSH model has at least $1$ and at most $N-1$ topological phases.
	
 \begin{figure*}[htb] %[t]
		\centering
		\includegraphics[max width=\linewidth]{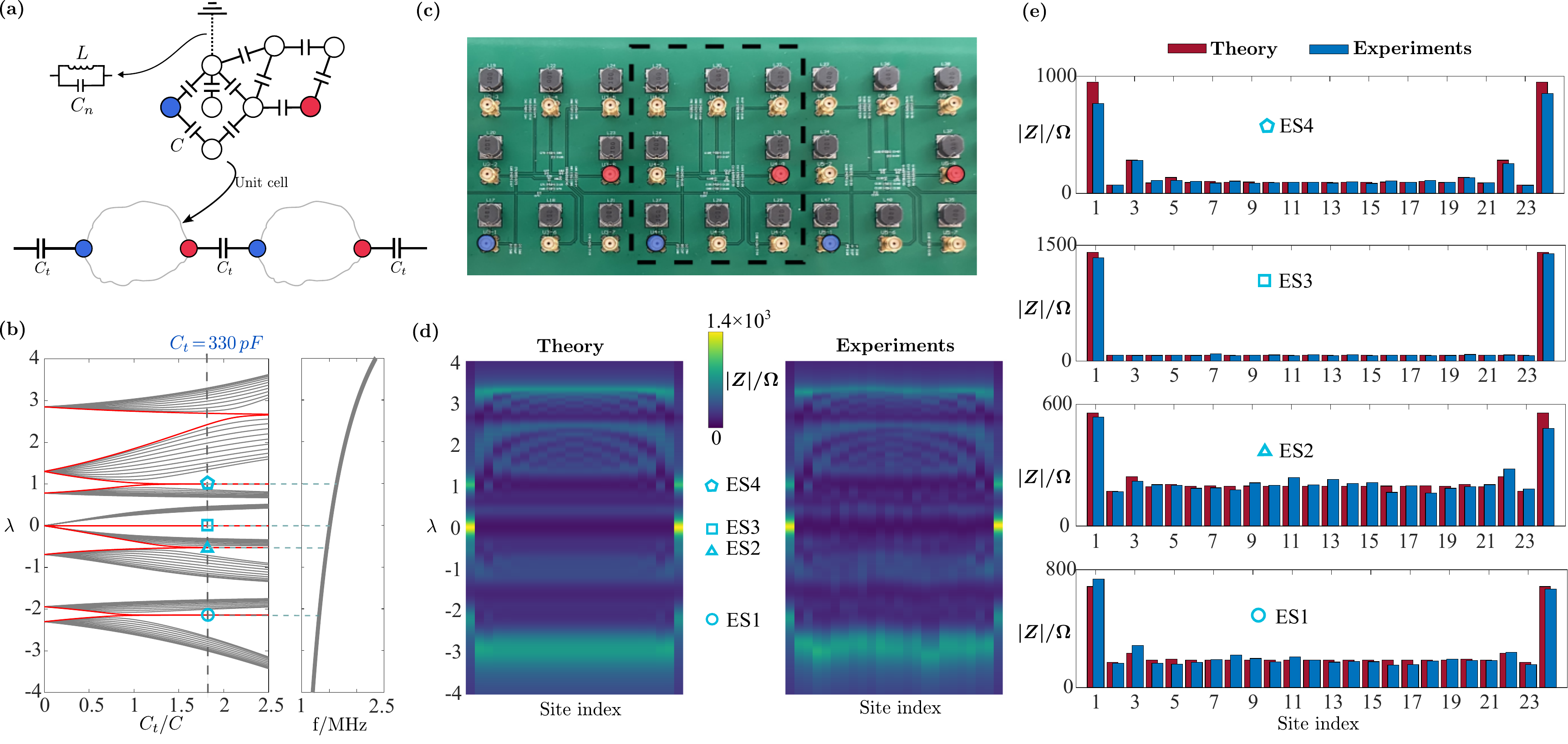}
		\caption{(a) Schematic of the latent topological circuit and its unit cell. Each site $n$ is grounded through a capacitor $C_n$ in parallel to an inductor $L$, though we depict this grounding only for one site to keep the figure as clear as possible. (b) Resonance frequency dispersion of a finite 12-unit-cell circuit in dependence of $C_t$. The topological edge states visible at $C_t = 330 pF$ are marked by four different geometrical shapes which are used in (d) and (e) as well.
        (c) Photograph showing three unit cells of the printed circuit.
        (d) Theoretical and experimental impedance value across all the latently symmetric points. (e) Impedance profile across the latently symmetric points at various frequencies that correspond to the topological edge states.}
		
		\label{fig:electricCircuitFigure}
	\end{figure*}
 \PRLSec{Experimental realization}
    In the following, we implement a latent SSH model with electrical circuits.
    We first review the theoretical description of the setup, which is depicted in \cref{fig:electricCircuitFigure} (a).
	Here, each site (vertex) of the original $8$-site setup of \cref{fig:figure1} (b) has been replaced by a junction, and each coupling (edge) by a capacitor. 
	Additionally, each site $n$ is now grounded through an inductance $L$ parallel to a capacitance $C_n = 5\,C-\Sigma_n$, where $\Sigma_n$ is the total capacitance of capacitors connecting site $n$ to other sites. 
	To link the electric circuit to the latent SSH model, we go to an eigenmode analysis.
    Starting from Kirchhoff's and Ohm's laws in frequency domain, it can be shown that the eigenmodes $\vecAlt{V}$ (whose entries describe the voltages at the junctions) and eigenfrequencies $\omega$ of the circuit are obtained by solving the eigenvalue problem \cite{Note17}
    \begin{equation} \label{eq:linEigProblem}
        H_P \vecAlt{V} = \lambda \vecAlt{V} \,,
    \end{equation}
    where $H_P$ is the tight-binding Hamiltonian of the $8$-site unit cell LS1 of \cref{fig:figure1} (b), and $\lambda = \frac{L \omega^2 (5C + C_t) - 1}{L C \omega^2}$.
    
    In order to test the predictions of the theory, we measured the circuit eigenmodes and eigenfrequencies, that is, their resonances.
    In circuit systems, these generally manifest themselves as peaks in the impedance spectrum, which can be conveniently measured \cite{Lee2018CP11TopolectricalCircuits,Imhof2018NP14925TopolectricalcircuitRealizationTopologicalCorner,Olekhno2020NC111436TopologicalEdgeStatesInteracting,Wang2020NC112356CircuitImplementationFourdimensionalTopological,Zhang2023NC141083HyperbolicBandTopologyNontrivial,Yang2023CP61RealizationWilsonFermionsTopolectrical,Wang2023PRL130057201RealizationHopfInsulatorCircuit,Li2020NC115491CriticalNonHermitianSkinEffect}.
    To see this, let us investigate the single-port impedance $Z_{a,a}$ of port $a$.
    This impedance can be obtained by connecting a cable to port $a$, applying a voltage $V_a$ and measuring the current $I_a$ into/out of this port; all other ports are not connected to external cables.
    Using Kirchhoff's and Ohm's law and some algebra, we see that $Z_{a,a}$ can be expanded in terms of the eigenvalues $\lambda_n$ and eigenvectors $\vecAlt{V}^{(n)}$ of $H_P$ as
        \begin{equation} \label{eq:ImpedanceExpansion}
        Z_{a,a}(\omega) = \frac{V_a}{I_a} =\frac{i}{\omega C}\left(\frac{1}{H_P - \lambda}\right)_{a,a} = \frac{i}{\omega C} \sum_n \frac{|V^{(n)}_{a}|^2}{\lambda_n - \lambda} \,.
    \end{equation}
    At resonance, that is, when the frequency $\omega$ is chosen such that $\lambda(\omega_R) \approx \lambda_n(\omega_R)$, the sum is dominated by the $n$-th term, and $Z_{a,a}(\omega_R)$ yields the spatial profile of $|\vecAlt{V}^{(n)}|^2$.
    We can thus determine \emph{both} the eigenfrequencies and eigenmodes of the circuit by measuring the frequency-dependent single-port impedances along the circuit.
    
    For our experiments, we fabricated a circuit comprising $12$ unit cells using standard printed circuit board (PCB) technique. A snapshot of three unit-cells used in our experiments is given in \cref{fig:electricCircuitFigure} (c).
    The substrate material is FR4 with 1Oz copper layer on both top and bottom layers. Low loss wire wound inductors (CY105-100K) are used, whose nominal inductance are $10uH$ with 10\% tolerance. However, our measurements show that the actual inductance are approximately $9uH$, with a parasitic resistance around $2\,\Omega$ at the frequency range of interest \cite{Note17}.
    Unless otherwise mentioned, the theoretical calculations are performed with these values \cite{Note17}.
    To allow for the impedance measurements, each node in our circuit is connected to a SMA port [cf. \cref{fig:electricCircuitFigure} (c)].
    Utilizing a vector network analyzer (R\&S ZNL3) \cite{Liu2020LSA9145OctupoleCornerStateThreedimensional,Liu2020PRA13014047GainLossInducedTopologicalInsulating}, we first measured the circuit's single-port scattering parameters $S(\omega)$, and subsequently converted them to impedance by $Z=Z_0\frac{1-S}{1+S}$, with $Z_0=50\Omega$ being the characteristic impedance of the microwave cable.

    In \cref{fig:electricCircuitFigure} (b), we show a theoretical computation of the eigenfrequency spectrum of our circuit in dependence of the inter-cell capacitance $C_t$, with $9uH$ grounding inductances but without parasitic resistance according to \cref{eq:linEigProblem}.
    Solid red lines represent the localized edge states \footnote{Note that the latently symmetric sites at the very boundary requires an extra $C_t$ grounding capacitor for faithfully corresponding to the tight-binding model; see \cite{Note17}.}.
    For our experiments shown in the following, we have chosen $C_t = 330\,pF$, for which the setup has four (out of the five possible) topological edge state pairs.
    Further experimental results for $C_t=820\, pF$, where all five topological edge state pairs are visible, are illustrated in the Supplemental Material \cite{Note17}.
    In \cref{fig:electricCircuitFigure} (d), we show both the theoretical computation and the measured values of the single-port impedance $Z$.
    Both quantities are shown as a two-dimensional color-map, with the free parameters being the operating frequency and site number.
    Topological edge states are found at the four frequencies ES1 to ES4, with the corresponding impedance measurements being shown in more detail in \cref{fig:electricCircuitFigure} (e).
    Overall, our experiments are in excellent agreement with the theory.
    We note that the deviation in localization behavior from the idealized situation (exponential decay) is due to the non-negligible parasite resistance of inductors, which is included in our theoretical calculations.
	
	\PRLSec{Conclusions and Outlook}
    In this work, we have demonstrated---both theoretically and experimentally---that one can use graph-theoretical principles to construct nonsymmetric systems whose isospectral reduction has the form of the SSH model.
    Furthermore, we showed that such latent SSH models are extensions of the original SSH-chain: They inherit the emergence of topological edge states, but are augmented by having many more than just a single topological transition.
    In other words, we combined the richness of a cavity with the topological nature of the SSH model.
    In the near future, this augmentation principle could be applied to other models---be they topological or non-topological, hermitian or non-hermitian---as well.
    A natural extension would be, for instance, to consider a latent non-hermitian SSH model.

	\begin{acknowledgments}
    \PRLSec{Acknowledgments}
		The authors are thankful to Olivier Richoux and Georgios Theocharis for valuable discussions.
	\end{acknowledgments}

%\bibliography{Bibtex.bib}

\begin{thebibliography}{43}%
	\makeatletter
	\providecommand \@ifxundefined [1]{%
		\@ifx{#1\undefined}
	}%
	\providecommand \@ifnum [1]{%
		\ifnum #1\expandafter \@firstoftwo
		\else \expandafter \@secondoftwo
		\fi
	}%
	\providecommand \@ifx [1]{%
		\ifx #1\expandafter \@firstoftwo
		\else \expandafter \@secondoftwo
		\fi
	}%
	\providecommand \natexlab [1]{#1}%
	\providecommand \enquote  [1]{``#1''}%
	\providecommand \bibnamefont  [1]{#1}%
	\providecommand \bibfnamefont [1]{#1}%
	\providecommand \citenamefont [1]{#1}%
	\providecommand \href@noop [0]{\@secondoftwo}%
	\providecommand \href [0]{\begingroup \@sanitize@url \@href}%
	\providecommand \@href[1]{\@@startlink{#1}\@@href}%
	\providecommand \@@href[1]{\endgroup#1\@@endlink}%
	\providecommand \@sanitize@url [0]{\catcode `\\12\catcode `\$12\catcode
		`\&12\catcode `\#12\catcode `\^12\catcode `\_12\catcode `\%12\relax}%
	\providecommand \@@startlink[1]{}%
	\providecommand \@@endlink[0]{}%
	\providecommand \url  [0]{\begingroup\@sanitize@url \@url }%
	\providecommand \@url [1]{\endgroup\@href {#1}{\urlprefix }}%
	\providecommand \urlprefix  [0]{URL }%
	\providecommand \Eprint [0]{\href }%
	\providecommand \doibase [0]{https://doi.org/}%
	\providecommand \selectlanguage [0]{\@gobble}%
	\providecommand \bibinfo  [0]{\@secondoftwo}%
	\providecommand \bibfield  [0]{\@secondoftwo}%
	\providecommand \translation [1]{[#1]}%
	\providecommand \BibitemOpen [0]{}%
	\providecommand \bibitemStop [0]{}%
	\providecommand \bibitemNoStop [0]{.\EOS\space}%
	\providecommand \EOS [0]{\spacefactor3000\relax}%
	\providecommand \BibitemShut  [1]{\csname bibitem#1\endcsname}%
	\let\auto@bib@innerbib\@empty
	%</preamble>
	\bibitem [{\citenamefont {Su}\ \emph {et~al.}(1979)\citenamefont {Su},
		\citenamefont {Schrieffer},\ and\ \citenamefont
		{Heeger}}]{Su1979PRL421698SolitonsPolyacetylene}%
	\BibitemOpen
	\bibfield  {author} {\bibinfo {author} {\bibfnamefont {W.~P.}\ \bibnamefont
			{Su}}, \bibinfo {author} {\bibfnamefont {J.~R.}\ \bibnamefont {Schrieffer}},\
		and\ \bibinfo {author} {\bibfnamefont {A.~J.}\ \bibnamefont {Heeger}},\
	}\bibfield  {title} {\bibinfo {title} {Solitons in polyacetylene},\ }\href
	{https://doi.org/10.1103/PhysRevLett.42.1698} {\bibfield  {journal} {\bibinfo
			{journal} {Phys. Rev. Lett.}\ }\textbf {\bibinfo {volume} {42}},\ \bibinfo
		{pages} {1698} (\bibinfo {year} {1979})}\BibitemShut {NoStop}%
	\bibitem [{\citenamefont {Asb{\'o}th}\ \emph {et~al.}(2016)\citenamefont
		{Asb{\'o}th}, \citenamefont {Oroszl{\'a}ny},\ and\ \citenamefont
		{P{\'a}lyi}}]{Asboth2016ShortCourseTopologicalInsulators}%
	\BibitemOpen
	\bibfield  {author} {\bibinfo {author} {\bibfnamefont {J.~K.}\ \bibnamefont
			{Asb{\'o}th}}, \bibinfo {author} {\bibfnamefont {L.}~\bibnamefont
			{Oroszl{\'a}ny}},\ and\ \bibinfo {author} {\bibfnamefont {A.}~\bibnamefont
			{P{\'a}lyi}},\ }\href@noop {} {\emph {\bibinfo {title} {A short course on
				topological insulators: band structure and edge states in one and two
				dimension}}},\ \bibinfo {edition} {1st}\ ed.,\ \bibinfo {series} {Lecture
		{{Notes}} in {{Physics}}}\ No.\ \bibinfo {number} {919}\ (\bibinfo
	{publisher} {{Springer International Publishing}},\ \bibinfo {address}
	{{Basel, Switzerland}},\ \bibinfo {year} {2016})\BibitemShut {NoStop}%
	\bibitem [{\citenamefont {Chen}\ \emph {et~al.}(2020)\citenamefont {Chen},
		\citenamefont {Wang}, \citenamefont {Zhang},\ and\ \citenamefont
		{Ma}}]{Chen2020PRA14024023ChiralSymmetryBreakingTightBinding}%
	\BibitemOpen
	\bibfield  {author} {\bibinfo {author} {\bibfnamefont {Z.-G.}\ \bibnamefont
			{Chen}}, \bibinfo {author} {\bibfnamefont {L.}~\bibnamefont {Wang}}, \bibinfo
		{author} {\bibfnamefont {G.}~\bibnamefont {Zhang}},\ and\ \bibinfo {author}
		{\bibfnamefont {G.}~\bibnamefont {Ma}},\ }\bibfield  {title} {\bibinfo
		{title} {Chiral symmetry breaking of tight-binding models in coupled
			acoustic-cavity systems},\ }\href
	{https://doi.org/10.1103/PhysRevApplied.14.024023} {\bibfield  {journal}
		{\bibinfo  {journal} {Phys. Rev. Appl.}\ }\textbf {\bibinfo {volume} {14}},\
		\bibinfo {pages} {024023} (\bibinfo {year} {2020})}\BibitemShut {NoStop}%
	\bibitem [{\citenamefont {Coutant}\ \emph {et~al.}(2021)\citenamefont
		{Coutant}, \citenamefont {Sivadon}, \citenamefont {Zheng}, \citenamefont
		{Achilleos}, \citenamefont {Richoux}, \citenamefont {Theocharis},\ and\
		\citenamefont
		{Pagneux}}]{Coutant2021PRB103224309AcousticSuSchriefferHeegerLatticeDirect}%
	\BibitemOpen
	\bibfield  {author} {\bibinfo {author} {\bibfnamefont {A.}~\bibnamefont
			{Coutant}}, \bibinfo {author} {\bibfnamefont {A.}~\bibnamefont {Sivadon}},
		\bibinfo {author} {\bibfnamefont {L.}~\bibnamefont {Zheng}}, \bibinfo
		{author} {\bibfnamefont {V.}~\bibnamefont {Achilleos}}, \bibinfo {author}
		{\bibfnamefont {O.}~\bibnamefont {Richoux}}, \bibinfo {author} {\bibfnamefont
			{G.}~\bibnamefont {Theocharis}},\ and\ \bibinfo {author} {\bibfnamefont
			{V.}~\bibnamefont {Pagneux}},\ }\bibfield  {title} {\bibinfo {title}
		{Acoustic {{Su-Schrieffer-Heeger}} lattice: {{Direct}} mapping of acoustic
			waveguides to the {{Su-Schrieffer-Heeger}} model},\ }\href
	{https://doi.org/10.1103/PhysRevB.103.224309} {\bibfield  {journal} {\bibinfo
			{journal} {Phys. Rev. B}\ }\textbf {\bibinfo {volume} {103}},\ \bibinfo
		{pages} {224309} (\bibinfo {year} {2021})}\BibitemShut {NoStop}%
	\bibitem [{\citenamefont {Kremer}\ \emph {et~al.}(2021)\citenamefont {Kremer},
		\citenamefont {Maczewsky}, \citenamefont {Heinrich},\ and\ \citenamefont
		{Szameit}}]{Kremer2021OMEO111014TopologicalEffectsIntegratedPhotonic}%
	\BibitemOpen
	\bibfield  {author} {\bibinfo {author} {\bibfnamefont {M.}~\bibnamefont
			{Kremer}}, \bibinfo {author} {\bibfnamefont {L.~J.}\ \bibnamefont
			{Maczewsky}}, \bibinfo {author} {\bibfnamefont {M.}~\bibnamefont
			{Heinrich}},\ and\ \bibinfo {author} {\bibfnamefont {A.}~\bibnamefont
			{Szameit}},\ }\bibfield  {title} {\bibinfo {title} {Topological effects in
			integrated photonic waveguide structures},\ }\href
	{https://doi.org/10.1364/OME.414648} {\bibfield  {journal} {\bibinfo
			{journal} {Opt. Mater. Express}\ }\textbf {\bibinfo {volume} {11}},\ \bibinfo
		{pages} {1014} (\bibinfo {year} {2021})}\BibitemShut {NoStop}%
	\bibitem [{\citenamefont {Tian}\ \emph {et~al.}(2019)\citenamefont {Tian},
		\citenamefont {Ke}, \citenamefont {Zhang}, \citenamefont {Lin}, \citenamefont
		{Shi}, \citenamefont {Huang}, \citenamefont {Lee},\ and\ \citenamefont
		{Du}}]{Tian2019PRB100024310ObservationDynamicalPhaseTransitions}%
	\BibitemOpen
	\bibfield  {author} {\bibinfo {author} {\bibfnamefont {T.}~\bibnamefont
			{Tian}}, \bibinfo {author} {\bibfnamefont {Y.}~\bibnamefont {Ke}}, \bibinfo
		{author} {\bibfnamefont {L.}~\bibnamefont {Zhang}}, \bibinfo {author}
		{\bibfnamefont {S.}~\bibnamefont {Lin}}, \bibinfo {author} {\bibfnamefont
			{Z.}~\bibnamefont {Shi}}, \bibinfo {author} {\bibfnamefont {P.}~\bibnamefont
			{Huang}}, \bibinfo {author} {\bibfnamefont {C.}~\bibnamefont {Lee}},\ and\
		\bibinfo {author} {\bibfnamefont {J.}~\bibnamefont {Du}},\ }\bibfield
	{title} {\bibinfo {title} {Observation of dynamical phase transitions in a
			topological nanomechanical system},\ }\href
	{https://doi.org/10.1103/PhysRevB.100.024310} {\bibfield  {journal} {\bibinfo
			{journal} {Phys. Rev. B}\ }\textbf {\bibinfo {volume} {100}},\ \bibinfo
		{pages} {024310} (\bibinfo {year} {2019})}\BibitemShut {NoStop}%
	\bibitem [{\citenamefont {Lee}\ \emph {et~al.}(2018)\citenamefont {Lee},
		\citenamefont {Imhof}, \citenamefont {Berger}, \citenamefont {Bayer},
		\citenamefont {Brehm}, \citenamefont {Molenkamp}, \citenamefont {Kiessling},\
		and\ \citenamefont {Thomale}}]{Lee2018CP11TopolectricalCircuits}%
	\BibitemOpen
	\bibfield  {author} {\bibinfo {author} {\bibfnamefont {C.~H.}\ \bibnamefont
			{Lee}}, \bibinfo {author} {\bibfnamefont {S.}~\bibnamefont {Imhof}}, \bibinfo
		{author} {\bibfnamefont {C.}~\bibnamefont {Berger}}, \bibinfo {author}
		{\bibfnamefont {F.}~\bibnamefont {Bayer}}, \bibinfo {author} {\bibfnamefont
			{J.}~\bibnamefont {Brehm}}, \bibinfo {author} {\bibfnamefont {L.~W.}\
			\bibnamefont {Molenkamp}}, \bibinfo {author} {\bibfnamefont {T.}~\bibnamefont
			{Kiessling}},\ and\ \bibinfo {author} {\bibfnamefont {R.}~\bibnamefont
			{Thomale}},\ }\bibfield  {title} {\bibinfo {title} {Topolectrical circuits},\
	}\href {https://doi.org/10.1038/s42005-018-0035-2} {\bibfield  {journal}
		{\bibinfo  {journal} {Commun. Phys.}\ }\textbf {\bibinfo {volume} {1}},\
		\bibinfo {pages} {1} (\bibinfo {year} {2018})}\BibitemShut {NoStop}%
	\bibitem [{\citenamefont {Guo}\ and\ \citenamefont
		{Chen}(2015)}]{Guo2015PRB91041402KaleidoscopeSymmetryprotectedTopologicalPhases}%
	\BibitemOpen
	\bibfield  {author} {\bibinfo {author} {\bibfnamefont {H.}~\bibnamefont
			{Guo}}\ and\ \bibinfo {author} {\bibfnamefont {S.}~\bibnamefont {Chen}},\
	}\bibfield  {title} {\bibinfo {title} {Kaleidoscope of symmetry-protected
			topological phases in one-dimensional periodically modulated lattices},\
	}\href {https://doi.org/10.1103/PhysRevB.91.041402} {\bibfield  {journal}
		{\bibinfo  {journal} {Phys. Rev. B}\ }\textbf {\bibinfo {volume} {91}},\
		\bibinfo {pages} {041402} (\bibinfo {year} {2015})}\BibitemShut {NoStop}%
	\bibitem [{\citenamefont {Anastasiadis}\ \emph {et~al.}(2022)\citenamefont
		{Anastasiadis}, \citenamefont {Styliaris}, \citenamefont {Chaunsali},
		\citenamefont {Theocharis},\ and\ \citenamefont
		{Diakonos}}]{Anastasiadis2022PRB106085109BulkedgeCorrespondenceTrimerSuSchriefferHeeger}%
	\BibitemOpen
	\bibfield  {author} {\bibinfo {author} {\bibfnamefont {A.}~\bibnamefont
			{Anastasiadis}}, \bibinfo {author} {\bibfnamefont {G.}~\bibnamefont
			{Styliaris}}, \bibinfo {author} {\bibfnamefont {R.}~\bibnamefont
			{Chaunsali}}, \bibinfo {author} {\bibfnamefont {G.}~\bibnamefont
			{Theocharis}},\ and\ \bibinfo {author} {\bibfnamefont {F.~K.}\ \bibnamefont
			{Diakonos}},\ }\bibfield  {title} {\bibinfo {title} {Bulk-edge correspondence
			in the trimer {{Su-Schrieffer-Heeger}} model},\ }\href
	{https://doi.org/10.1103/PhysRevB.106.085109} {\bibfield  {journal} {\bibinfo
			{journal} {Phys. Rev. B}\ }\textbf {\bibinfo {volume} {106}},\ \bibinfo
		{pages} {085109} (\bibinfo {year} {2022})}\BibitemShut {NoStop}%
	\bibitem [{\citenamefont {Xie}\ \emph {et~al.}(2019)\citenamefont {Xie},
		\citenamefont {Gou}, \citenamefont {Xiao}, \citenamefont {Gadway},\ and\
		\citenamefont {Yan}}]{Xie2019nQI51TopologicalCharacterizationsExtendedSu}%
	\BibitemOpen
	\bibfield  {author} {\bibinfo {author} {\bibfnamefont {D.}~\bibnamefont
			{Xie}}, \bibinfo {author} {\bibfnamefont {W.}~\bibnamefont {Gou}}, \bibinfo
		{author} {\bibfnamefont {T.}~\bibnamefont {Xiao}}, \bibinfo {author}
		{\bibfnamefont {B.}~\bibnamefont {Gadway}},\ and\ \bibinfo {author}
		{\bibfnamefont {B.}~\bibnamefont {Yan}},\ }\bibfield  {title} {\bibinfo
		{title} {Topological characterizations of an extended
			{{Su}}\textendash{{Schrieffer}}\textendash{{Heeger}} model},\ }\href
	{https://doi.org/10.1038/s41534-019-0159-6} {\bibfield  {journal} {\bibinfo
			{journal} {npj Quantum Inf.}\ }\textbf {\bibinfo {volume} {5}},\ \bibinfo
		{pages} {1} (\bibinfo {year} {2019})}\BibitemShut {NoStop}%
	\bibitem [{\citenamefont {{P{\'e}rez-Gonz{\'a}lez}}\ \emph
		{et~al.}(2019)\citenamefont {{P{\'e}rez-Gonz{\'a}lez}}, \citenamefont
		{Bello}, \citenamefont {{G{\'o}mez-Le{\'o}n}},\ and\ \citenamefont
		{Platero}}]{Perez-Gonzalez2019PRB99035146InterplayLongrangeHoppingDisorder}%
	\BibitemOpen
	\bibfield  {author} {\bibinfo {author} {\bibfnamefont {B.}~\bibnamefont
			{{P{\'e}rez-Gonz{\'a}lez}}}, \bibinfo {author} {\bibfnamefont
			{M.}~\bibnamefont {Bello}}, \bibinfo {author} {\bibfnamefont
			{{\'A}.}~\bibnamefont {{G{\'o}mez-Le{\'o}n}}},\ and\ \bibinfo {author}
		{\bibfnamefont {G.}~\bibnamefont {Platero}},\ }\bibfield  {title} {\bibinfo
		{title} {Interplay between long-range hopping and disorder in topological
			systems},\ }\href {https://doi.org/10.1103/PhysRevB.99.035146} {\bibfield
		{journal} {\bibinfo  {journal} {Phys. Rev. B}\ }\textbf {\bibinfo {volume}
			{99}},\ \bibinfo {pages} {035146} (\bibinfo {year} {2019})}\BibitemShut
	{NoStop}%
	\bibitem [{\citenamefont {Biswas}\ \emph {et~al.}(2022)\citenamefont {Biswas},
		\citenamefont {Mukherjee},\ and\ \citenamefont
		{Chakrabarti}}]{Biswas2022FlatBandsEdgeStates}%
	\BibitemOpen
	\bibfield  {author} {\bibinfo {author} {\bibfnamefont {S.}~\bibnamefont
			{Biswas}}, \bibinfo {author} {\bibfnamefont {A.}~\bibnamefont {Mukherjee}},\
		and\ \bibinfo {author} {\bibfnamefont {A.}~\bibnamefont {Chakrabarti}},\
	}\href@noop {} {\bibinfo {title} {Flat bands, edge states and possible
			topological phases in a branching fractal}} (\bibinfo {year} {2022}),\
	\Eprint {https://arxiv.org/abs/2209.05117} {arxiv:2209.05117 [cond-mat]}
	\BibitemShut {NoStop}%
	\bibitem [{\citenamefont {Smith}\ and\ \citenamefont
		{Webb}(2019)}]{Smith2019PA514855HiddenSymmetriesRealTheoretical}%
	\BibitemOpen
	\bibfield  {author} {\bibinfo {author} {\bibfnamefont {D.}~\bibnamefont
			{Smith}}\ and\ \bibinfo {author} {\bibfnamefont {B.}~\bibnamefont {Webb}},\
	}\bibfield  {title} {\bibinfo {title} {Hidden symmetries in real and
			theoretical networks},\ }\href {https://doi.org/10.1016/j.physa.2018.09.131}
	{\bibfield  {journal} {\bibinfo  {journal} {Physica A}\ }\textbf {\bibinfo
			{volume} {514}},\ \bibinfo {pages} {855} (\bibinfo {year}
		{2019})}\BibitemShut {NoStop}%
	\bibitem [{\citenamefont {Morfonios}\ \emph
		{et~al.}(2021{\natexlab{a}})\citenamefont {Morfonios}, \citenamefont
		{R{\"o}ntgen}, \citenamefont {Pyzh},\ and\ \citenamefont
		{Schmelcher}}]{Morfonios2021PRB104035105FlatBandsLatentSymmetry}%
	\BibitemOpen
	\bibfield  {author} {\bibinfo {author} {\bibfnamefont {C.~V.}\ \bibnamefont
			{Morfonios}}, \bibinfo {author} {\bibfnamefont {M.}~\bibnamefont
			{R{\"o}ntgen}}, \bibinfo {author} {\bibfnamefont {M.}~\bibnamefont {Pyzh}},\
		and\ \bibinfo {author} {\bibfnamefont {P.}~\bibnamefont {Schmelcher}},\
	}\bibfield  {title} {\bibinfo {title} {Flat bands by latent symmetry},\
	}\href {https://doi.org/10.1103/PhysRevB.104.035105} {\bibfield  {journal}
		{\bibinfo  {journal} {Phys. Rev. B}\ }\textbf {\bibinfo {volume} {104}},\
		\bibinfo {pages} {035105} (\bibinfo {year} {2021}{\natexlab{a}})}\BibitemShut
	{NoStop}%
	\bibitem [{\citenamefont {Leykam}\ \emph {et~al.}(2018)\citenamefont {Leykam},
		\citenamefont {Andreanov},\ and\ \citenamefont
		{Flach}}]{Leykam2018AP31473052ArtificialFlatBandSystems}%
	\BibitemOpen
	\bibfield  {author} {\bibinfo {author} {\bibfnamefont {D.}~\bibnamefont
			{Leykam}}, \bibinfo {author} {\bibfnamefont {A.}~\bibnamefont {Andreanov}},\
		and\ \bibinfo {author} {\bibfnamefont {S.}~\bibnamefont {Flach}},\ }\bibfield
	{title} {\bibinfo {title} {Artificial flat band systems: from lattice models
			to experiments},\ }\href {https://doi.org/10.1080/23746149.2018.1473052}
	{\bibfield  {journal} {\bibinfo  {journal} {Adv. Phys.}\ }\textbf {\bibinfo
			{volume} {3}},\ \bibinfo {pages} {1473052} (\bibinfo {year}
		{2018})}\BibitemShut {NoStop}%
	\bibitem [{\citenamefont {R{\"o}ntgen}\ \emph {et~al.}(2021)\citenamefont
		{R{\"o}ntgen}, \citenamefont {Pyzh}, \citenamefont {Morfonios}, \citenamefont
		{Palaiodimopoulos}, \citenamefont {Diakonos},\ and\ \citenamefont
		{Schmelcher}}]{Rontgen2021PRL126180601LatentSymmetryInducedDegeneracies}%
	\BibitemOpen
	\bibfield  {author} {\bibinfo {author} {\bibfnamefont {M.}~\bibnamefont
			{R{\"o}ntgen}}, \bibinfo {author} {\bibfnamefont {M.}~\bibnamefont {Pyzh}},
		\bibinfo {author} {\bibfnamefont {C.~V.}\ \bibnamefont {Morfonios}}, \bibinfo
		{author} {\bibfnamefont {N.~E.}\ \bibnamefont {Palaiodimopoulos}}, \bibinfo
		{author} {\bibfnamefont {F.~K.}\ \bibnamefont {Diakonos}},\ and\ \bibinfo
		{author} {\bibfnamefont {P.}~\bibnamefont {Schmelcher}},\ }\bibfield  {title}
	{\bibinfo {title} {Latent symmetry induced degeneracies},\ }\href
	{https://doi.org/10.1103/PhysRevLett.126.180601} {\bibfield  {journal}
		{\bibinfo  {journal} {Phys. Rev. Lett.}\ }\textbf {\bibinfo {volume} {126}},\
		\bibinfo {pages} {180601} (\bibinfo {year} {2021})}\BibitemShut {NoStop}%
	\bibitem [{\citenamefont {R{\"o}ntgen}\ \emph
		{et~al.}(2023{\natexlab{a}})\citenamefont {R{\"o}ntgen}, \citenamefont
		{Morfonios}, \citenamefont {Schmelcher},\ and\ \citenamefont
		{Pagneux}}]{Rontgen2023PRL130077201HiddenSymmetriesAcousticWave}%
	\BibitemOpen
	\bibfield  {author} {\bibinfo {author} {\bibfnamefont {M.}~\bibnamefont
			{R{\"o}ntgen}}, \bibinfo {author} {\bibfnamefont {C.~V.}\ \bibnamefont
			{Morfonios}}, \bibinfo {author} {\bibfnamefont {P.}~\bibnamefont
			{Schmelcher}},\ and\ \bibinfo {author} {\bibfnamefont {V.}~\bibnamefont
			{Pagneux}},\ }\bibfield  {title} {\bibinfo {title} {Hidden symmetries in
			acoustic wave systems},\ }\href
	{https://doi.org/10.1103/PhysRevLett.130.077201} {\bibfield  {journal}
		{\bibinfo  {journal} {Phys. Rev. Lett.}\ }\textbf {\bibinfo {volume} {130}},\
		\bibinfo {pages} {077201} (\bibinfo {year} {2023}{\natexlab{a}})}\BibitemShut
	{NoStop}%
	\bibitem [{\citenamefont {Sol}\ \emph {et~al.}()\citenamefont {Sol},
		\citenamefont {R{\"o}ntgen},\ and\ \citenamefont {del
			Hougne}}]{Sol2023CovertScatteringControlMetamaterials}%
	\BibitemOpen
	\bibfield  {author} {\bibinfo {author} {\bibfnamefont {J.}~\bibnamefont
			{Sol}}, \bibinfo {author} {\bibfnamefont {M.}~\bibnamefont {R{\"o}ntgen}},\
		and\ \bibinfo {author} {\bibfnamefont {P.}~\bibnamefont {del Hougne}},\
	}\bibfield  {title} {\bibinfo {title} {Covert scattering control in
			metamaterials with non-locally encoded hidden symmetry},\ }\href
	{https://doi.org/10.1002/adma.202303891} {\bibfield  {journal} {\bibinfo
			{journal} {Adv. Mater.}\ }\textbf {\bibinfo {volume} {In press}},\ \bibinfo
		{pages} {2303891}}\BibitemShut {NoStop}%
	\bibitem [{\citenamefont {R{\"o}ntgen}\ \emph
		{et~al.}(2023{\natexlab{b}})\citenamefont {R{\"o}ntgen}, \citenamefont
		{Richoux}, \citenamefont {Theocharis}, \citenamefont {Morfonios},
		\citenamefont {Schmelcher}, \citenamefont {{del Hougne}},\ and\ \citenamefont
		{Achilleos}}]{Rontgen2023EquireflectionalityCustomizedUnbalancedCoherent}%
	\BibitemOpen
	\bibfield  {author} {\bibinfo {author} {\bibfnamefont {M.}~\bibnamefont
			{R{\"o}ntgen}}, \bibinfo {author} {\bibfnamefont {O.}~\bibnamefont
			{Richoux}}, \bibinfo {author} {\bibfnamefont {G.}~\bibnamefont {Theocharis}},
		\bibinfo {author} {\bibfnamefont {C.~V.}\ \bibnamefont {Morfonios}}, \bibinfo
		{author} {\bibfnamefont {P.}~\bibnamefont {Schmelcher}}, \bibinfo {author}
		{\bibfnamefont {P.}~\bibnamefont {{del Hougne}}},\ and\ \bibinfo {author}
		{\bibfnamefont {V.}~\bibnamefont {Achilleos}},\ }\href
	{https://doi.org/10.48550/arXiv.2305.02786} {\bibinfo {title}
		{Equireflectionality and customized unbalanced coherent perfect absorption in
			asymmetric waveguide networks}} (\bibinfo {year} {2023}{\natexlab{b}}),\
	\Eprint {https://arxiv.org/abs/2305.02786} {arxiv:2305.02786 [physics]}
	\BibitemShut {NoStop}%
	\bibitem [{\citenamefont
		{R{\"o}ntgen}(2022)}]{Rontgen2022LatentSymmetriesIntroduction}%
	\BibitemOpen
	\bibfield  {author} {\bibinfo {author} {\bibfnamefont {M.}~\bibnamefont
			{R{\"o}ntgen}},\ }\href {https://doi.org/10.52843/meta-mat.7d7p2h} {\emph
		{\bibinfo {title} {Latent symmetries: {{An}} introduction}}},\ \bibinfo
	{type} {other}\ (\bibinfo  {institution} {{MetaMAT Weekly Seminars}},\
	\bibinfo {year} {2022})\BibitemShut {NoStop}%
	\bibitem [{Note17()}]{Note17}%
	\BibitemOpen
	\bibinfo {note} {See the Supplemental Material for more examples of latently
		symmetric setups, a proof that the Zak phase in latent SSH models is
		quantized, as well as details regarding the electric circuit
		part.}\BibitemShut {Stop}%
	\bibitem [{\citenamefont {Grosso}\ and\ \citenamefont
		{Parravicini}(2013)}]{Grosso2013SolidStatePhysics}%
	\BibitemOpen
	\bibfield  {author} {\bibinfo {author} {\bibfnamefont {G.}~\bibnamefont
			{Grosso}}\ and\ \bibinfo {author} {\bibfnamefont {G.~P.}\ \bibnamefont
			{Parravicini}},\ }\href@noop {} {\emph {\bibinfo {title} {Solid state
				physics}}}\ (\bibinfo  {publisher} {{Academic Press}},\ \bibinfo {year}
	{2013})\BibitemShut {NoStop}%
	\bibitem [{Note18()}]{Note18}%
	\BibitemOpen
	\bibinfo {note} {The eigenvalue spectrum of $\protect \cc@accent
		{"707E}{H}_S(E)$ is defined as the solutions to $Det(\protect \cc@accent
		{"707E}{H}_S(E) - E I )$.}\BibitemShut {Stop}%
	\bibitem [{\citenamefont {Bunimovich}\ and\ \citenamefont
		{Webb}(2014)}]{Bunimovich2014IsospectralTransformationsNewApproach}%
	\BibitemOpen
	\bibfield  {author} {\bibinfo {author} {\bibfnamefont {L.}~\bibnamefont
			{Bunimovich}}\ and\ \bibinfo {author} {\bibfnamefont {B.}~\bibnamefont
			{Webb}},\ }\href@noop {} {\emph {\bibinfo {title} {Isospectral
				transformations: a new approach to analyzing multidimensional systems and
				networks}}},\ \bibinfo {edition} {1st}\ ed.,\ Springer {{Monographs}} in
	{{Mathematics}}\ (\bibinfo  {publisher} {{Springer}},\ \bibinfo {address}
	{{New York, NY, United States}},\ \bibinfo {year} {2014})\BibitemShut
	{NoStop}%
	\bibitem [{Note19()}]{Note19}%
	\BibitemOpen
	\bibinfo {note} {Or when choosing $S$ such that it does not include sites
		which are related/mapped to each other by either a classical or a latent
		reflection symmetry.}\BibitemShut {Stop}%
	\bibitem [{Note20()}]{Note20}%
	\BibitemOpen
	\bibinfo {note} {This expression follows by analogy from the corresponding
		expression of the classic SSH model which can be found in chapter 1.5.6. of
		Ref. \cite {Asboth2016ShortCourseTopologicalInsulators}.}\BibitemShut {Stop}%
	\bibitem [{\citenamefont {Godsil}\ and\ \citenamefont
		{McKay}(1982)}]{Godsil1982AM25257ConstructingCospectralGraphs}%
	\BibitemOpen
	\bibfield  {author} {\bibinfo {author} {\bibfnamefont {C.~D.}\ \bibnamefont
			{Godsil}}\ and\ \bibinfo {author} {\bibfnamefont {B.~D.}\ \bibnamefont
			{McKay}},\ }\bibfield  {title} {\bibinfo {title} {Constructing cospectral
			graphs},\ }\href {https://doi.org/10.1007/BF02189621} {\bibfield  {journal}
		{\bibinfo  {journal} {Aequ. Math.}\ }\textbf {\bibinfo {volume} {25}},\
		\bibinfo {pages} {257} (\bibinfo {year} {1982})}\BibitemShut {NoStop}%
	\bibitem [{\citenamefont {McKay}\ and\ \citenamefont
		{Piperno}(2014)}]{McKay2014JoSC6094PracticalGraphIsomorphismII}%
	\BibitemOpen
	\bibfield  {author} {\bibinfo {author} {\bibfnamefont {B.~D.}\ \bibnamefont
			{McKay}}\ and\ \bibinfo {author} {\bibfnamefont {A.}~\bibnamefont
			{Piperno}},\ }\bibfield  {title} {\bibinfo {title} {Practical graph
			isomorphism, {{II}}},\ }\href {https://doi.org/10.1016/j.jsc.2013.09.003}
	{\bibfield  {journal} {\bibinfo  {journal} {J. Symb. Comput.}\ }\textbf
		{\bibinfo {volume} {60}},\ \bibinfo {pages} {94} (\bibinfo {year}
		{2014})}\BibitemShut {NoStop}%
	\bibitem [{\citenamefont {R{\"o}ntgen}\ \emph {et~al.}(2020)\citenamefont
		{R{\"o}ntgen}, \citenamefont {Palaiodimopoulos}, \citenamefont {Morfonios},
		\citenamefont {Brouzos}, \citenamefont {Pyzh}, \citenamefont {Diakonos},\
		and\ \citenamefont
		{Schmelcher}}]{Rontgen2020PRA101042304DesigningPrettyGoodState}%
	\BibitemOpen
	\bibfield  {author} {\bibinfo {author} {\bibfnamefont {M.}~\bibnamefont
			{R{\"o}ntgen}}, \bibinfo {author} {\bibfnamefont {N.~E.}\ \bibnamefont
			{Palaiodimopoulos}}, \bibinfo {author} {\bibfnamefont {C.~V.}\ \bibnamefont
			{Morfonios}}, \bibinfo {author} {\bibfnamefont {I.}~\bibnamefont {Brouzos}},
		\bibinfo {author} {\bibfnamefont {M.}~\bibnamefont {Pyzh}}, \bibinfo {author}
		{\bibfnamefont {F.~K.}\ \bibnamefont {Diakonos}},\ and\ \bibinfo {author}
		{\bibfnamefont {P.}~\bibnamefont {Schmelcher}},\ }\bibfield  {title}
	{\bibinfo {title} {Designing pretty good state transfer via isospectral
			reductions},\ }\href {https://doi.org/10.1103/PhysRevA.101.042304} {\bibfield
		{journal} {\bibinfo  {journal} {Phys. Rev. A}\ }\textbf {\bibinfo {volume}
			{101}},\ \bibinfo {pages} {042304} (\bibinfo {year} {2020})}\BibitemShut
	{NoStop}%
	\bibitem [{\citenamefont {Morfonios}\ \emph
		{et~al.}(2021{\natexlab{b}})\citenamefont {Morfonios}, \citenamefont {Pyzh},
		\citenamefont {R{\"o}ntgen},\ and\ \citenamefont
		{Schmelcher}}]{Morfonios2021LAaiA62453CospectralityPreservingGraphModifications}%
	\BibitemOpen
	\bibfield  {author} {\bibinfo {author} {\bibfnamefont {C.~V.}\ \bibnamefont
			{Morfonios}}, \bibinfo {author} {\bibfnamefont {M.}~\bibnamefont {Pyzh}},
		\bibinfo {author} {\bibfnamefont {M.}~\bibnamefont {R{\"o}ntgen}},\ and\
		\bibinfo {author} {\bibfnamefont {P.}~\bibnamefont {Schmelcher}},\ }\bibfield
	{title} {\bibinfo {title} {Cospectrality preserving graph modifications and
			eigenvector properties via walk equivalence of vertices},\ }\href
	{https://doi.org/10.1016/j.laa.2021.04.004} {\bibfield  {journal} {\bibinfo
			{journal} {Linear Algebra Its Appl.}\ }\textbf {\bibinfo {volume} {624}},\
		\bibinfo {pages} {53} (\bibinfo {year} {2021}{\natexlab{b}})}\BibitemShut
	{NoStop}%
	\bibitem [{\citenamefont {Imhof}\ \emph {et~al.}(2018)\citenamefont {Imhof},
		\citenamefont {Berger}, \citenamefont {Bayer}, \citenamefont {Brehm},
		\citenamefont {Molenkamp}, \citenamefont {Kiessling}, \citenamefont
		{Schindler}, \citenamefont {Lee}, \citenamefont {Greiter}, \citenamefont
		{Neupert},\ and\ \citenamefont
		{Thomale}}]{Imhof2018NP14925TopolectricalcircuitRealizationTopologicalCorner}%
	\BibitemOpen
	\bibfield  {author} {\bibinfo {author} {\bibfnamefont {S.}~\bibnamefont
			{Imhof}}, \bibinfo {author} {\bibfnamefont {C.}~\bibnamefont {Berger}},
		\bibinfo {author} {\bibfnamefont {F.}~\bibnamefont {Bayer}}, \bibinfo
		{author} {\bibfnamefont {J.}~\bibnamefont {Brehm}}, \bibinfo {author}
		{\bibfnamefont {L.~W.}\ \bibnamefont {Molenkamp}}, \bibinfo {author}
		{\bibfnamefont {T.}~\bibnamefont {Kiessling}}, \bibinfo {author}
		{\bibfnamefont {F.}~\bibnamefont {Schindler}}, \bibinfo {author}
		{\bibfnamefont {C.~H.}\ \bibnamefont {Lee}}, \bibinfo {author} {\bibfnamefont
			{M.}~\bibnamefont {Greiter}}, \bibinfo {author} {\bibfnamefont
			{T.}~\bibnamefont {Neupert}},\ and\ \bibinfo {author} {\bibfnamefont
			{R.}~\bibnamefont {Thomale}},\ }\bibfield  {title} {\bibinfo {title}
		{Topolectrical-circuit realization of topological corner modes},\ }\href
	{https://doi.org/10.1038/s41567-018-0246-1} {\bibfield  {journal} {\bibinfo
			{journal} {Nat. Phys.}\ }\textbf {\bibinfo {volume} {14}},\ \bibinfo
		{pages} {925} (\bibinfo {year} {2018})}\BibitemShut {NoStop}%
	\bibitem [{\citenamefont {Olekhno}\ \emph {et~al.}(2020)\citenamefont
		{Olekhno}, \citenamefont {Kretov}, \citenamefont {Stepanenko}, \citenamefont
		{Ivanova}, \citenamefont {Yaroshenko}, \citenamefont {Puhtina}, \citenamefont
		{Filonov}, \citenamefont {Cappello}, \citenamefont {Matekovits},\ and\
		\citenamefont
		{Gorlach}}]{Olekhno2020NC111436TopologicalEdgeStatesInteracting}%
	\BibitemOpen
	\bibfield  {author} {\bibinfo {author} {\bibfnamefont {N.~A.}\ \bibnamefont
			{Olekhno}}, \bibinfo {author} {\bibfnamefont {E.~I.}\ \bibnamefont {Kretov}},
		\bibinfo {author} {\bibfnamefont {A.~A.}\ \bibnamefont {Stepanenko}},
		\bibinfo {author} {\bibfnamefont {P.~A.}\ \bibnamefont {Ivanova}}, \bibinfo
		{author} {\bibfnamefont {V.~V.}\ \bibnamefont {Yaroshenko}}, \bibinfo
		{author} {\bibfnamefont {E.~M.}\ \bibnamefont {Puhtina}}, \bibinfo {author}
		{\bibfnamefont {D.~S.}\ \bibnamefont {Filonov}}, \bibinfo {author}
		{\bibfnamefont {B.}~\bibnamefont {Cappello}}, \bibinfo {author}
		{\bibfnamefont {L.}~\bibnamefont {Matekovits}},\ and\ \bibinfo {author}
		{\bibfnamefont {M.~A.}\ \bibnamefont {Gorlach}},\ }\bibfield  {title}
	{\bibinfo {title} {Topological edge states of interacting photon pairs
			emulated in a topolectrical circuit},\ }\href
	{https://doi.org/10.1038/s41467-020-14994-7} {\bibfield  {journal} {\bibinfo
			{journal} {Nat. Commun.}\ }\textbf {\bibinfo {volume} {11}},\ \bibinfo
		{pages} {1436} (\bibinfo {year} {2020})}\BibitemShut {NoStop}%
	\bibitem [{\citenamefont {Wang}\ \emph {et~al.}(2020)\citenamefont {Wang},
		\citenamefont {Price}, \citenamefont {Zhang},\ and\ \citenamefont
		{Chong}}]{Wang2020NC112356CircuitImplementationFourdimensionalTopological}%
	\BibitemOpen
	\bibfield  {author} {\bibinfo {author} {\bibfnamefont {Y.}~\bibnamefont
			{Wang}}, \bibinfo {author} {\bibfnamefont {H.~M.}\ \bibnamefont {Price}},
		\bibinfo {author} {\bibfnamefont {B.}~\bibnamefont {Zhang}},\ and\ \bibinfo
		{author} {\bibfnamefont {Y.~D.}\ \bibnamefont {Chong}},\ }\bibfield  {title}
	{\bibinfo {title} {Circuit implementation of a four-dimensional topological
			insulator},\ }\href {https://doi.org/10.1038/s41467-020-15940-3} {\bibfield
		{journal} {\bibinfo  {journal} {Nat. Commun.}\ }\textbf {\bibinfo {volume}
			{11}},\ \bibinfo {pages} {2356} (\bibinfo {year} {2020})}\BibitemShut
	{NoStop}%
	\bibitem [{\citenamefont {Zhang}\ \emph {et~al.}(2023)\citenamefont {Zhang},
		\citenamefont {Di}, \citenamefont {Zheng}, \citenamefont {Sun},\ and\
		\citenamefont {Zhang}}]{Zhang2023NC141083HyperbolicBandTopologyNontrivial}%
	\BibitemOpen
	\bibfield  {author} {\bibinfo {author} {\bibfnamefont {W.}~\bibnamefont
			{Zhang}}, \bibinfo {author} {\bibfnamefont {F.}~\bibnamefont {Di}}, \bibinfo
		{author} {\bibfnamefont {X.}~\bibnamefont {Zheng}}, \bibinfo {author}
		{\bibfnamefont {H.}~\bibnamefont {Sun}},\ and\ \bibinfo {author}
		{\bibfnamefont {X.}~\bibnamefont {Zhang}},\ }\bibfield  {title} {\bibinfo
		{title} {Hyperbolic band topology with non-trivial second {{Chern}}
			numbers},\ }\href {https://doi.org/10.1038/s41467-023-36767-8} {\bibfield
		{journal} {\bibinfo  {journal} {Nat. Commun.}\ }\textbf {\bibinfo {volume}
			{14}},\ \bibinfo {pages} {1083} (\bibinfo {year} {2023})}\BibitemShut
	{NoStop}%
	\bibitem [{\citenamefont {Yang}\ \emph {et~al.}(2023)\citenamefont {Yang},
		\citenamefont {Song}, \citenamefont {Cao},\ and\ \citenamefont
		{Yan}}]{Yang2023CP61RealizationWilsonFermionsTopolectrical}%
	\BibitemOpen
	\bibfield  {author} {\bibinfo {author} {\bibfnamefont {H.}~\bibnamefont
			{Yang}}, \bibinfo {author} {\bibfnamefont {L.}~\bibnamefont {Song}}, \bibinfo
		{author} {\bibfnamefont {Y.}~\bibnamefont {Cao}},\ and\ \bibinfo {author}
		{\bibfnamefont {P.}~\bibnamefont {Yan}},\ }\bibfield  {title} {\bibinfo
		{title} {Realization of {{Wilson}} fermions in topolectrical circuits},\
	}\href {https://doi.org/10.1038/s42005-023-01326-6} {\bibfield  {journal}
		{\bibinfo  {journal} {Commun. Phys.}\ }\textbf {\bibinfo {volume} {6}},\
		\bibinfo {pages} {1} (\bibinfo {year} {2023})}\BibitemShut {NoStop}%
	\bibitem [{\citenamefont {Wang}\ \emph {et~al.}(2023)\citenamefont {Wang},
		\citenamefont {Zeng}, \citenamefont {Biao}, \citenamefont {Yan},\ and\
		\citenamefont {Yu}}]{Wang2023PRL130057201RealizationHopfInsulatorCircuit}%
	\BibitemOpen
	\bibfield  {author} {\bibinfo {author} {\bibfnamefont {Z.}~\bibnamefont
			{Wang}}, \bibinfo {author} {\bibfnamefont {X.-T.}\ \bibnamefont {Zeng}},
		\bibinfo {author} {\bibfnamefont {Y.}~\bibnamefont {Biao}}, \bibinfo {author}
		{\bibfnamefont {Z.}~\bibnamefont {Yan}},\ and\ \bibinfo {author}
		{\bibfnamefont {R.}~\bibnamefont {Yu}},\ }\bibfield  {title} {\bibinfo
		{title} {Realization of a {{Hopf}} insulator in circuit systems},\ }\href
	{https://doi.org/10.1103/PhysRevLett.130.057201} {\bibfield  {journal}
		{\bibinfo  {journal} {Phys. Rev. Lett.}\ }\textbf {\bibinfo {volume} {130}},\
		\bibinfo {pages} {057201} (\bibinfo {year} {2023})}\BibitemShut {NoStop}%
	\bibitem [{\citenamefont {Li}\ \emph {et~al.}(2020)\citenamefont {Li},
		\citenamefont {Lee}, \citenamefont {Mu},\ and\ \citenamefont
		{Gong}}]{Li2020NC115491CriticalNonHermitianSkinEffect}%
	\BibitemOpen
	\bibfield  {author} {\bibinfo {author} {\bibfnamefont {L.}~\bibnamefont
			{Li}}, \bibinfo {author} {\bibfnamefont {C.~H.}\ \bibnamefont {Lee}},
		\bibinfo {author} {\bibfnamefont {S.}~\bibnamefont {Mu}},\ and\ \bibinfo
		{author} {\bibfnamefont {J.}~\bibnamefont {Gong}},\ }\bibfield  {title}
	{\bibinfo {title} {Critical non-{{Hermitian}} skin effect},\ }\href
	{https://doi.org/10.1038/s41467-020-18917-4} {\bibfield  {journal} {\bibinfo
			{journal} {Nat. Commun.}\ }\textbf {\bibinfo {volume} {11}},\ \bibinfo
		{pages} {5491} (\bibinfo {year} {2020})}\BibitemShut {NoStop}%
	\bibitem [{\citenamefont {Liu}\ \emph {et~al.}(2020{\natexlab{a}})\citenamefont
		{Liu}, \citenamefont {Ma}, \citenamefont {Zhang}, \citenamefont {Zhang},
		\citenamefont {Yang}, \citenamefont {You}, \citenamefont {Gao}, \citenamefont
		{Xiang}, \citenamefont {Cui},\ and\ \citenamefont
		{Zhang}}]{Liu2020LSA9145OctupoleCornerStateThreedimensional}%
	\BibitemOpen
	\bibfield  {author} {\bibinfo {author} {\bibfnamefont {S.}~\bibnamefont
			{Liu}}, \bibinfo {author} {\bibfnamefont {S.}~\bibnamefont {Ma}}, \bibinfo
		{author} {\bibfnamefont {Q.}~\bibnamefont {Zhang}}, \bibinfo {author}
		{\bibfnamefont {L.}~\bibnamefont {Zhang}}, \bibinfo {author} {\bibfnamefont
			{C.}~\bibnamefont {Yang}}, \bibinfo {author} {\bibfnamefont {O.}~\bibnamefont
			{You}}, \bibinfo {author} {\bibfnamefont {W.}~\bibnamefont {Gao}}, \bibinfo
		{author} {\bibfnamefont {Y.}~\bibnamefont {Xiang}}, \bibinfo {author}
		{\bibfnamefont {T.~J.}\ \bibnamefont {Cui}},\ and\ \bibinfo {author}
		{\bibfnamefont {S.}~\bibnamefont {Zhang}},\ }\bibfield  {title} {\bibinfo
		{title} {Octupole corner state in a three-dimensional topological circuit},\
	}\href {https://doi.org/10.1038/s41377-020-00381-w} {\bibfield  {journal}
		{\bibinfo  {journal} {Light Sci. Appl.}\ }\textbf {\bibinfo {volume} {9}},\
		\bibinfo {pages} {145} (\bibinfo {year} {2020}{\natexlab{a}})}\BibitemShut
	{NoStop}%
	\bibitem [{\citenamefont {Liu}\ \emph {et~al.}(2020{\natexlab{b}})\citenamefont
		{Liu}, \citenamefont {Ma}, \citenamefont {Yang}, \citenamefont {Zhang},
		\citenamefont {Gao}, \citenamefont {Xiang}, \citenamefont {Cui},\ and\
		\citenamefont
		{Zhang}}]{Liu2020PRA13014047GainLossInducedTopologicalInsulating}%
	\BibitemOpen
	\bibfield  {author} {\bibinfo {author} {\bibfnamefont {S.}~\bibnamefont
			{Liu}}, \bibinfo {author} {\bibfnamefont {S.}~\bibnamefont {Ma}}, \bibinfo
		{author} {\bibfnamefont {C.}~\bibnamefont {Yang}}, \bibinfo {author}
		{\bibfnamefont {L.}~\bibnamefont {Zhang}}, \bibinfo {author} {\bibfnamefont
			{W.}~\bibnamefont {Gao}}, \bibinfo {author} {\bibfnamefont {Y.~J.}\
			\bibnamefont {Xiang}}, \bibinfo {author} {\bibfnamefont {T.~J.}\ \bibnamefont
			{Cui}},\ and\ \bibinfo {author} {\bibfnamefont {S.}~\bibnamefont {Zhang}},\
	}\bibfield  {title} {\bibinfo {title} {Gain- and loss-induced topological
			insulating phase in a non-hermitian electrical circuit},\ }\href
	{https://doi.org/10.1103/PhysRevApplied.13.014047} {\bibfield  {journal}
		{\bibinfo  {journal} {Phys. Rev. Applied}\ }\textbf {\bibinfo {volume}
			{13}},\ \bibinfo {pages} {014047} (\bibinfo {year}
		{2020}{\natexlab{b}})}\BibitemShut {NoStop}%
	\bibitem [{Note21()}]{Note21}%
	\BibitemOpen
	\bibinfo {note} {Note that the latently symmetric sites at the very boundary
		requires an extra $C_t$ grounding capacitor for faithfully corresponding to
		the tight-binding model; see \cite {Note17}.}\BibitemShut {Stop}%
	\bibitem [{\citenamefont {Kempton}\ \emph {et~al.}(2020)\citenamefont
		{Kempton}, \citenamefont {Sinkovic}, \citenamefont {Smith},\ and\
		\citenamefont
		{Webb}}]{Kempton2020LAIA594226CharacterizingCospectralVerticesIsospectral}%
	\BibitemOpen
	\bibfield  {author} {\bibinfo {author} {\bibfnamefont {M.}~\bibnamefont
			{Kempton}}, \bibinfo {author} {\bibfnamefont {J.}~\bibnamefont {Sinkovic}},
		\bibinfo {author} {\bibfnamefont {D.}~\bibnamefont {Smith}},\ and\ \bibinfo
		{author} {\bibfnamefont {B.}~\bibnamefont {Webb}},\ }\bibfield  {title}
	{\bibinfo {title} {Characterizing cospectral vertices via isospectral
			reduction},\ }\href {https://doi.org/10.1016/j.laa.2020.02.020} {\bibfield
		{journal} {\bibinfo  {journal} {Linear Algebra Its Appl.}\ }\textbf {\bibinfo
			{volume} {594}},\ \bibinfo {pages} {226} (\bibinfo {year}
		{2020})}\BibitemShut {NoStop}%
	\bibitem [{\citenamefont {Godsil}\ and\ \citenamefont
		{Smith}(2017)}]{Godsil2017A1StronglyCospectralVertices}%
	\BibitemOpen
	\bibfield  {author} {\bibinfo {author} {\bibfnamefont {C.}~\bibnamefont
			{Godsil}}\ and\ \bibinfo {author} {\bibfnamefont {J.}~\bibnamefont {Smith}},\
	}\href {https://doi.org/10.48550/arXiv.1709.07975} {\bibinfo {title}
		{Strongly cospectral vertices}} (\bibinfo {year} {2017}),\ \Eprint
	{https://arxiv.org/abs/1709.07975} {arxiv:1709.07975} \BibitemShut {NoStop}%
	\bibitem [{Note22()}]{Note22}%
	\BibitemOpen
	\bibinfo {note} {In this Lemma, the phrasing ``latently symmetric'' is not
		used; instead, the two sites $u,v$ are said to be cospectral. For a
		real-symmetric matrix $H$, cospectrality of two sites $u,v$ and the statement
		that they are latently symmetric are equivalent; see Ref. \cite
		{Kempton2020LAIA594226CharacterizingCospectralVerticesIsospectral}}\BibitemShut
	{NoStop}%
\end{thebibliography}

\onecolumngrid
\appendix
\clearpage

\section{More examples of latently symmetric setups}
\label{app:LSExamples}
In the main text, we mentioned that one can easily generate a plethora of latently symmetric unit cells by relying on graph-theoretical tools.
\Cref{fig:zakPhase} (b) to (d) show three example systems that have been obtained with this method.
We further mentioned that one can analyze the matrix powers of a given latently symmetric setup to derive changes to the system that preserve its latent symmetry.
In \cref{fig:zakPhase} (e), we visualize the outcome of this approach---whose derivation can be found in \cite{Morfonios2021LAaiA62453CospectralityPreservingGraphModifications}---for the unit cell LS2 from Fig. 1 (b) of the main text.
The latent symmetry of sites $u,v$ is preserved for any choice of the real coupling strengths $a,b,c,d,e,f$.
Moreover, one may freely and individually vary the on-site potentials of sites $x,y,z$, and---though by keeping these two identical---the on-site potentials of sites $u,v$.

\section{Quantization of the Zak phase due to latent symmetry}
\label{app:Zak}

     \begin{figure}[htb] %[t]
	\centering
	\includegraphics[max width=\linewidth]{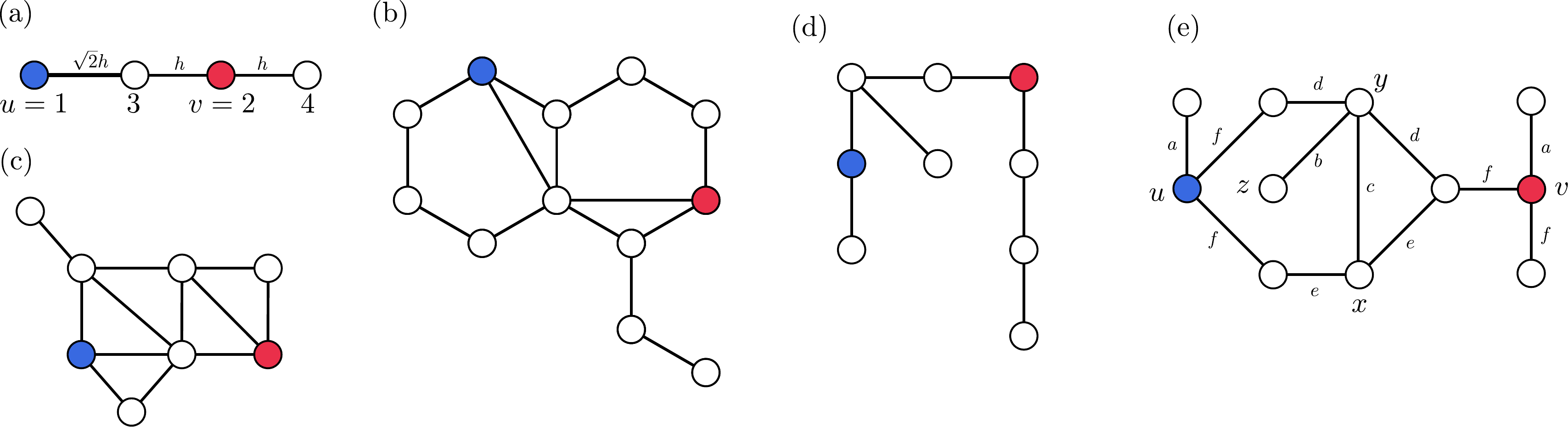}
	\caption{Five latently symmetric systems (see text for details). Unless otherwise marked, all couplings are equal to unity.
	}
	\label{fig:zakPhase}
\end{figure}

In the following, we show that a latent SSH model features a quantized Zak phase.
To this end, let us start by assuming that our unit cell Hamiltonian is real-valued and with two latently symmetric sites $u,v$, like the one depicted in \cref{fig:zakPhase}. 
To be explicit, let us choose the setup depicted in \cref{fig:zakPhase} (a), whose Hamiltonian reads
\begin{equation}
    \ham = h \left(
\begin{array}{cccc}
 0 & 0 & \sqrt{2} & 0 \\
 0 & 0 & 1 & 1 \\
 \sqrt{2} & 1 & 0 & 0 \\
 0 & 1 & 0 & 0 \\
\end{array} \, .
\right)
\end{equation}
Now, whenever we have such a real-valued latently symmetric Hamiltonian, there exists a block-diagonal matrix (numbering the sites such that $S=\{u,v\}$ are the first two)
\begin{equation}
    Q = \begin{pmatrix}
        0 & 1 \\
        1 & 0 \\
        & & \overline{Q}
    \end{pmatrix}
\end{equation}
which (i) commutes with $\ham$, (ii) is symmetric, that is, $Q = Q^{T}$, (iii) is orthogonal, $Q^{T} = Q^{-1}$ [see Lemma 11.1  in Ref. \cite{Godsil2017A1StronglyCospectralVertices},  \footnote{In this Lemma, the phrasing ``latently symmetric'' is not used; instead, the two sites $u,v$ are said to be cospectral. For a real-symmetric matrix $\ham$, cospectrality of two sites $u,v$ and the statement that they are latently symmetric are equivalent; see Ref. \cite{Kempton2020LAIA594226CharacterizingCospectralVerticesIsospectral}}, or Sec. IV. of the Supplemental Material of Ref. \cite{Rontgen2021PRL126180601LatentSymmetryInducedDegeneracies}], from which it follows that $Q^2 = \idMat$.
In other words, $Q$ acts as a permutation on the two sites $S$, while it acts as an orthogonal transformation $\overline{Q}$ on the sites $\sbar$.
For our example of \cref{fig:zakPhase}, we have
\begin{equation}
    Q = \left(
\begin{array}{cccc}
 0 & 1 & 0 & 0 \\
 1 & 0 & 0 & 0 \\
 0 & 0 & \frac{1}{\sqrt{2}} & \frac{1}{\sqrt{2}} \\
 0 & 0 & \frac{1}{\sqrt{2}} & -\frac{1}{\sqrt{2}} \\
\end{array}
\right) \, .
\end{equation}

Now, when we use our latently symmetric Hamiltonian $\ham$ as a unit cell and build a lattice by connecting sites $u,v$ of neighboring unit cells, our Bloch Hamiltonian $\ham_B(k)$ fulfills
\begin{equation} \label{Mirror_Bloch_H}
Q \cdot \ham_B(k) \cdot Q = \ham_B(-k) \,.
\end{equation}
This means in particular that if $\varphi_n(k)$ is an eigenvector for $\ham_B(k)$, $Q\cdot \varphi_n(k)$ is an eigenvector for $\ham_B(-k)$ with the same eigenvalue. Because we assume the eigenvalues to be all non-degenerate (non overlapping bands) this means 
\begin{equation}
\varphi_n(-k) = e^{i \theta(k)} M_x \cdot \varphi_n(k), 
\end{equation}
where $\theta(k)$ is a locally smooth function of $k$. In particular, on the band edges, $k=0$ or $k = \pi$ are mirror symmetric momenta (since $k=-\pi$ is equivalent to $k=\pi$). At these values of $k$, $Q$ commutes with the Hamiltonian, and hence, $\varphi_n$ is either symmetric or antisymmetric, i.e. $\theta(0)$ and $\theta(\pi)$ are either $0$ or $\pi$. 

If $\varphi_n(q)$ is the Bloch eigenvector of the $n^{\rm th}$ band, we define the Berry connection \cite{Asboth2016ShortCourseTopologicalInsulators}
\begin{equation}
    A_n(k) = -i \braket{\varphi_n | \partial_k \varphi_n} \,.
\end{equation}
The Zak phase is defined as the integral of the Berry connection over the Brillouin zone: 
\begin{equation} \label{ZakPhase_def}
\alpha_Z = \oint_{-\pi}^\pi A_n(k) \d k. 
\end{equation}

The mirror symmetry commutation relation with the Hamiltonian in equation \cref{Mirror_Bloch_H} gives us a relation between the Berry connection at $k$ and $-k$.
After taking the derivative of \cref{Mirror_Bloch_H} and taking the scalar product with $\varphi_n(-k)$, we see that 
\begin{equation} 
A_n(-k) = A_n(k) + \partial_q \theta . 
\end{equation}  
In other words, the Berry connection at $k$ differs from that at $-k$ by a gauge. Now, if we integrate that relation over half of the Brillouin zone, we obtain 
\begin{equation} 
-\int_{-\pi}^0 A_n(k) \d q = \int_{0}^{\pi} A_n(k) \d q + \theta(\pi) - \theta(0). 
\end{equation} 
Combining the two integrals gives the integral over the whole Brillouin zone, and hence, the Zak phase. The latter then has the simple expression 
\begin{equation} \label{1DZak_Sym}
\alpha_Z = \theta(\pi) - \theta(0)\,,  
\end{equation} 
from which it follows that either $\alpha_Z = 0 \textrm{ mod } 2\pi$ (both edge eigenvectors have the same symmetry) or $\alpha_Z = \pi \textrm{ mod } 2\pi$ (the two eigenvectors do not have the same symmetry).

\section{Details regarding the experimental setup} \label{app:expDetails}

\begin{figure}[H] %[t]
	\centering
	\includegraphics[max width=1\linewidth]{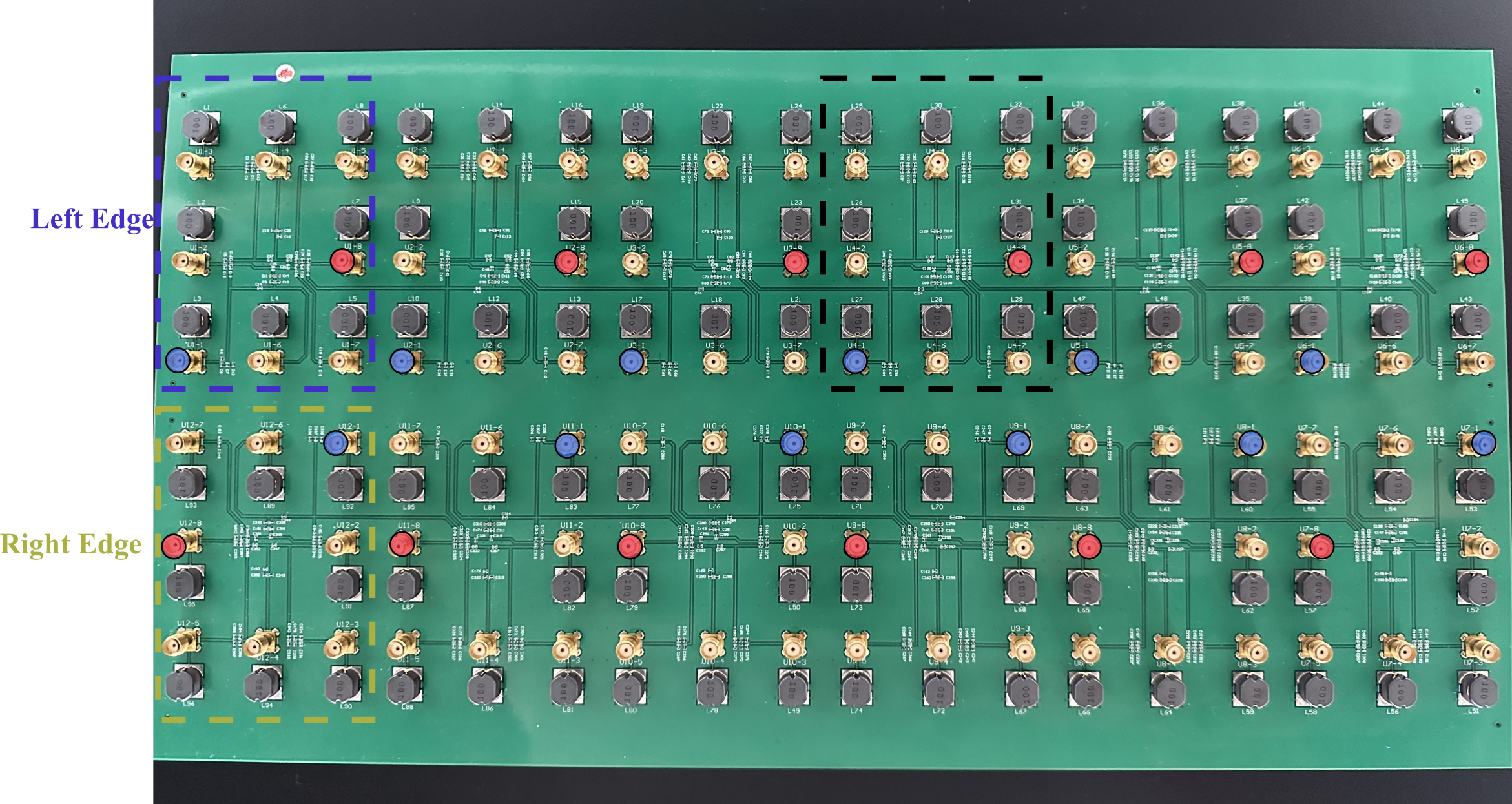}
	\caption{Snapshot of the full 12 unit-cell latent SSH circuit.
	}
	\label{fig:full_circuit}
\end{figure}

Before going into details, let us start by providing a snapshot of the full 12 unit-cell latent SSH circuits; cf. \cref{fig:full_circuit}. The latently symmetric points are marked by blue/red circles. Unit-cells 6 to 12 are folded back to keep the circuit compact. Left most and right most cells are highlighted by blue and yellow dashed lines, respectively. Co-spectral sites are highlighted by blue and red full circles, as they are in \cref{fig:electricCircuitFigure}(b).

\subsection{Deriving the linear eigenvalue problem and the impedance formula}
In the following, we shall derive the linear eigenvalue problem of \cref{eq:linEigProblem}.

To do so, we assume that each node/site of the circuit is connected via an extra cable to the exterior, so that current can flow into/out of the circuit. Denoting by $\vecAlt{V}$ the voltage at the current's nodes, and by $\vecAlt{I}$ the current flowing into/out of the nodes, we can use Kirchhoff's and Ohm's law in frequency space to derive the important relation
\begin{equation} \label{eq:groundedLaplacian}
    \jMat(\omega)\vecAlt{V} = \vecAlt{I} \,.
\end{equation}
Here, $\jMat(\omega)=-i\omega CH_P+i\omega (5\,C+C_t)\idMat+\frac{1}{i\omega L}\idMat$ is the so-called circuit Laplacian \cite{Lee2018CP11TopolectricalCircuits} with $H_P$ the tight-binding Hamiltonian of the $8$-site model of \cref{fig:figure1} (b).
Eigenmodes occur when there is no external current flowing into the circuit, that is, $\vecAlt{I} = 0$, and we obtain the quadratic eigenvalue problem
    \begin{equation} \label{eq:zeroMode}
        \jMat(\omega)\vecAlt{V} = 0 \,.
    \end{equation}
    Here,  .
    Since the second and third term of $\jMat$ are proportional to the identity matrix, \cref{eq:zeroMode} can be written as the linear eigenvalue problem $\ham \vecAlt{V} = \frac{1}{\omega^2} \vecAlt{V}$
	with $\ham = L \left( (5C + C_{t}) \idMat - C H_P\right)$.
    \Cref{eq:linEigProblem} is then a direct consequence.
    We note that we also have
    \begin{equation} \label{eq:jMatInHP}
        \jMat(\omega) = -i \omega C \left( H_P - \lambda \right) \,.
    \end{equation}
    When calculating the single-port impedance in \cref{eq:ImpedanceExpansion}, the relations \cref{eq:groundedLaplacian,eq:jMatInHP} can be used to express
    \begin{equation}\label{eq:ImpedanceExpansion1}
        Z_{a,a}(\omega) = \frac{V_a}{I_a} = \left(J^{-1}\right)_{a,a} = \frac{i}{\omega C}\left( \left(H_P - \lambda \right)^{-1}\right)_{a,a}\,,
    \end{equation}
    which is the expression used in \cref{eq:ImpedanceExpansion}.
    When parasitic resistances $R$ of the inductors are considered, the circuit Laplacian is modified as 
    \begin{equation} \label{eq:paraResistance}
        \jMat(\omega)=-i\omega CH_P+i\omega (5\,C+C_t)\idMat+\frac{1}{i\omega L+R}\idMat \,.
    \end{equation}
    In which $R$ is the parasitic serial resistance of the inductors (see below) and can be subsequently implemented for the single-port impedance calculations. Accordingly under parasitic resistance, \cref{eq:ImpedanceExpansion} can be modified as  
     \begin{equation} \label{eq:ImpedanceExpansion_lossy}
      \begin{split}
        Z_{a,a}(\omega) &= \frac{i}{\omega C} \sum_n \frac{|V^{(n)}_{a}|^2}{\lambda_n - \lambda+\eta} \, \\
        \eta&=-\frac{1}{\omega^2LC}\frac{R}{R+i\omega L} \,.
      \end{split}
     \end{equation}

\subsection{Inductance/parasite resistance measurements}
\begin{figure}[htb] %[t]
	\centering
	\includegraphics[max width=1\linewidth]{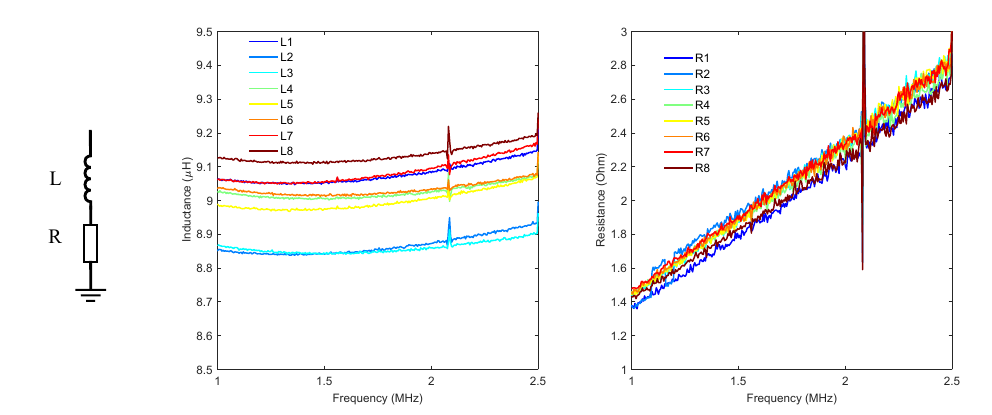}
	\caption{Experimentally measured inductance and the parasitic resistance of the inductors utilized in the latent SSH circuit
	}
	\label{fig:inductors}
\end{figure}

In the following, we show the experimentally measured inductance and parasite resistance of the inductors. The results are given in \cref{fig:inductors}. To do so, we unsolder all the grounding and coupling capacitors of the circuit and measure impedance of grounding inductors only. The impedance value are then converted to inductance and resistance for each frequency, assuming they are serially connected; see \cref{fig:inductors}. We found that the inductance are lower than the nominal value of $10 \,\mu H$ and are close to $9\,\mu H$. The parasite resistance dispersed almost linearly from 1.4 $Ohm$ to 2.8 $Ohm$, which are significantly larger than the nominal DC resistance, 0.034 $Ohm$, probably due to skin effect induced extra loss at higher frequencies. Here we omit discussing the detailed parasite effects in such surface-mounted components that introduce dispersed inductance and resistance rather than constants, and adopting constant values that are close to experiments in our theoretical model, granted such means has already provided accurate predictions and are consistent with experimental results.  

\subsection{Experiments with $C_t = 820 \, pF$}
\begin{figure}[htb] %[t]
	\centering
	\includegraphics[max width=0.9\linewidth]{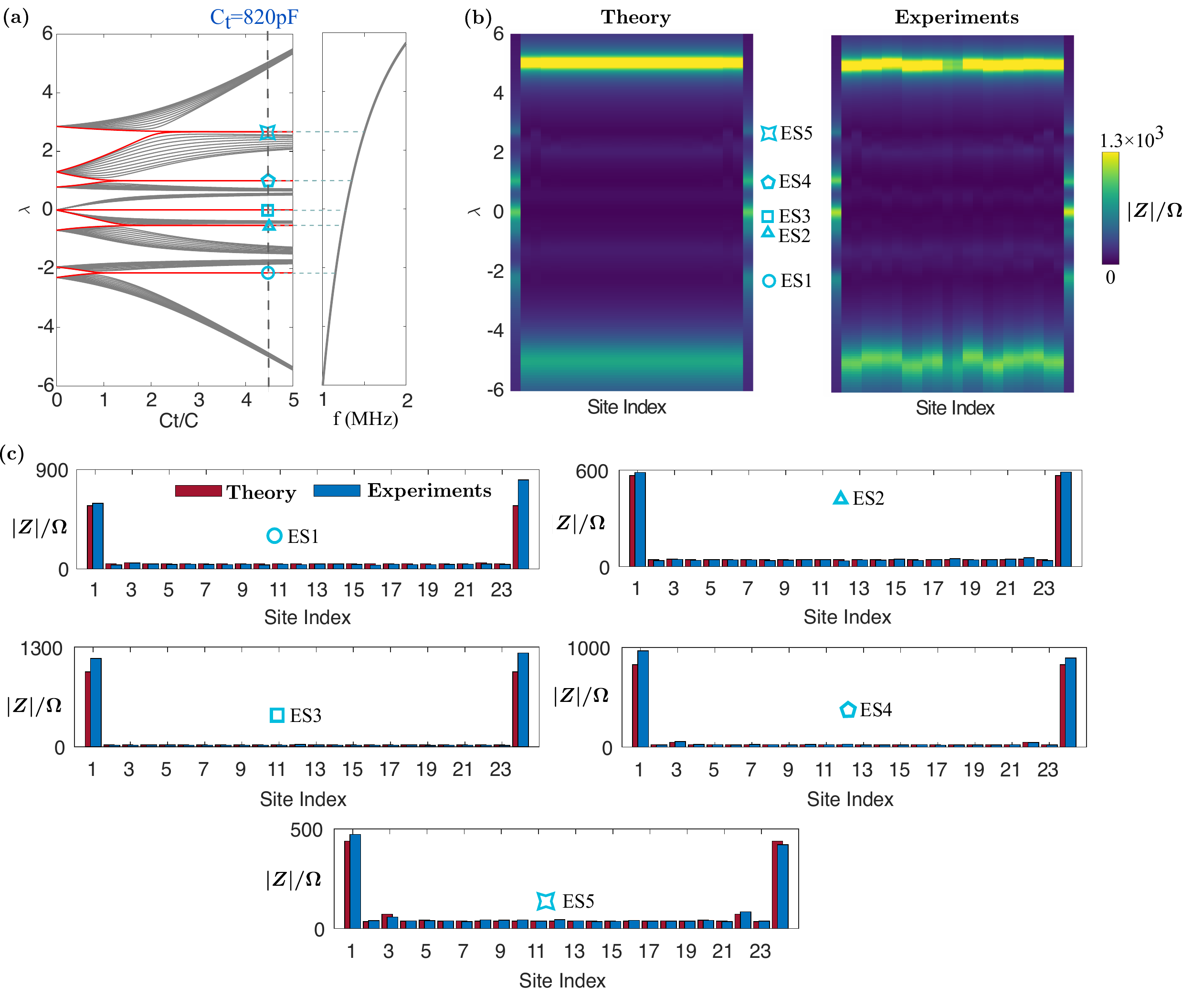}
	\caption{(a) Resonance frequency dispersion of a finite 12-unit-cell circuit in dependence of $C_t$. The topological edge states visible at $C_t = 820 pF$ are marked by five different geometrical shapes which are used in (b) and (c) as well. (b) Theoretical and experimental impedance value across all the latently symmetric points. (c) Impedance profile across the latently symmetric points at various frequencies that correspond to the topological edge states.
	}
	\label{fig:820pF}
\end{figure}

In \cref{fig:820pF}, we show the comparisons of experimental and theoretical results with $C_t=820\,pF$, for which the setup has five topological edge state pairs.
In \cref{fig:820pF} (a), we show a theoretical computation of the eigenfrequency spectrum of our circuit in dependence of the inter-cell capacitance $C_t$, with $9uH$ grounding inductances but without parasitic resistance, same as \cref{fig:electricCircuitFigure}. 
Solid red lines represent the localized edge states 
In \cref{fig:820pF} (b), we show both the theoretical computation and the measured values of the single-port impedance $Z$.
Both quantities are shown as a two-dimensional color-map, with the free parameters being the operating frequency and site number.
Topological edge states are found at the five frequencies ES1 to ES5, with the corresponding impedance measurements being shown in more detail in \cref{fig:820pF} (c). Overall, since the chosen coupling capacitance $C_t = 820\,pF$ is located deep in the topological region, localization of the edge states is stronger than in \cref{fig:electricCircuitFigure}.

\end{document}